%% file: main.tex
\documentclass[manuscript,screen,nonacm]{acmart}

\usepackage{textcomp}
\usepackage{xcolor}
\usepackage{algorithm}
\usepackage{float}
\usepackage{url}
\usepackage{siunitx}
\usepackage{multirow}
\usepackage{subcaption}
\usepackage{pifont}

\newlength{\headerlength}
\newcommand{\headerangle}{90}
\AtBeginDocument{%
  \providecommand\BibTeX{{%
    \normalfont B\kern-0.5em{\scshape i\kern-0.25em b}\kern-0.8em\TeX}}}

\setcopyright{acmcopyright}
\copyrightyear{2023}
\acmYear{2023}
\acmDOI{XXXXXXX.XXXXXXX}

\begin{document}

\title{Engineering End-to-End Remote Labs using IoT-based Retrofitting}

\author{K. S. Viswanadh}
\orcid{0009-0009-0969-0953}
\email{savitha.viswanadh@research.iiit.ac.in}
\author{A. Gureja}
\email{akshit.gureja@research.iiit.ac.in}
\orcid{0009-0001-8420-0100}
\author{N. Walchatwar}
\email{nagesh.walchatwar@research.iiit.ac.in}
\orcid{0009-0001-4286-2456}
\author{R. Agrawal}
\orcid{0009-0008-6205-0077}
\email{rishabh.agrawal@students.iiit.ac.in}

\author{S. Sinha}
\email{shiven.sinha@research.iiit.ac.in}
\orcid{0009-0000-5259-2683}

\author{S. Chaudhari}
\email{sachin.chaudhari@iiit.ac.in}
\orcid{0000-0003-1923-0925}

\author{K. Vaidhyanathan}
\email{karthik.vaidhyanathan@iiit.ac.in}
\orcid{0000-0003-2317-6175}

\author{V. Choppella}
\email{venkatesh.choppella@iiit.ac.in}
\orcid{0000-0003-1085-3464}

\author{P. Bhimalapuram}
\email{prabhakar.b@iiit.ac.in}
\orcid{0000-0002-8302-1696}

\author{H. Kandath}
\email{harikumar.k@iiit.ac.in}
\orcid{0000-0002-5500-7133}

\author{A. Hussain}
\email{aftab.hussain@iiit.ac.in}
\orcid{0000-0002-9516-9428}

\affiliation{%
  \institution{International Institution of Information Technology, Hyderabad}
  \city{Hyderabad}
  \state{Telangana}
  \country{India}
  \postcode{500032}
}

\renewcommand{\shortauthors}{K. S. Viswanadh, et al.}

\begin{abstract}
Remote labs are a groundbreaking development in the education industry, providing students with access to laboratory education anytime, anywhere. However, most remote labs are costly and difficult to scale, especially in developing countries. With this as a motivation, this paper proposes a new remote labs (RLabs) solution that includes two use case experiments: Vanishing Rod and Focal Length. The hardware experiments are built at a low-cost by retrofitting Internet of Things (IoT) components. They are also made portable by designing miniaturised and modular setups. The software architecture designed as part of the solution seamlessly supports the scalability of the experiments, offering compatibility with a wide range of hardware devices and IoT platforms. Additionally, it can live-stream remote experiments without needing dedicated server space for the stream. The software architecture also includes an automation suite that periodically checks the status of the experiments using computer vision (CV). RLabs is qualitatively evaluated against seven non-functional attributes - affordability, portability, scalability, compatibility, maintainability, usability, and universality. Finally, user feedback was collected from a group of students, and the scores indicate a positive response to the students’ learning and the platform’s usability.
\end{abstract}




\maketitle

\section{Introduction} \label{sec:Intro}
\input{source_intro}

\section{Use case Experiments} \label{sec:model}

\input{source_model}


\section{Hardware: Design and Implementation} \label{sec:hardware}
\input{source_hardware}


\section{Software: RLabs Platform, Implementation and Testing} \label{sec:software}
\input{source_software}


\section{Results}\label{sec:results}
\input{source_results}


\section{Conclusions} \label{sec:conc}

This study presents the development of a remote lab system named RLabs that included the development of two use case experiments along with a software platform. The proposed system is qualitatively evaluated against seven NFAs - affordability, portability, scalability, compatibility, maintainability, usability, and universality. The experiments were built by retrofitting IoT components on traditional laboratory equipment. Modular and miniaturised versions of the same experiments are also built using 3D-printed components. Miniaturised experiments are lower in cost by 41 \% and 24 \% for Vanishing Rod and Focal Length experiments compared to the retrofitted setups, making them even more affordable. Similarly, there is also a reduction in weight and volume by 5.4 and 12 times for the Vanishing Rod experiment and 2.8 and 4.4 times for the Focal Length experiment, respectively, showing the compactness and portability of the miniaturised setups. The system is scalable as many experiments can be built at low-cost and with fewer materials. At the same time, the architecture of the software platform allows many experiments to be hosted without consuming many resources due to the implementation of the P2P live-streaming service. The compatibility of the system is shown by connecting different hardware boards (like Raspberry Pi, ESP32), IoT platforms (like Blynk and Thingspeak) and cameras (RaspiCam, USB camera and IP Camera). Also, the platform is tested by operating it on different devices, web browsers and operating systems. An automated testing suite has monitored the experiments' operational status for four months and reported an uptime of 84 \%. The usability survey, filled out by a group of forty-five high school students, showed an average score of 4.34, indicating a positive learning experience and good usability of the system.


\begin{acks}
This work was funded in part by Enabler Grants made available through the Kohli Center for Intelligent Systems (KCIS) at IIIT Hyderabad, Raj Reddy Center for Technology and Society (RCTS) and TIH Foundation for IoT \& IoE at IIT Bombay under the CHANAKYA Fellowship Program 2022-23 (3) (Grant No: TIH-IoT/2023-03/HRD/CHANAKYA/SL/CFP-016), with no conflicts of interest. The authors thank Shikhar Educare (Amravati, Maharashtra, India) for providing valuable feedback on the RLabs system. 
\end{acks}

\bibliographystyle{ACM-Reference-Format}
\bibliography{references}

\end{document}

%% file: source_intro.tex
\subsection{Motivation}

Several nations across the globe face issues in maintaining the quality of education due to factors like inadequate funding, a shortage of qualified teachers, and a lack of well-equipped laboratory facilities, particularly in rural areas ~\cite{unesco2019, worldbank2018}. This situation is often exacerbated by external challenges such as conflicts, natural disasters, and health crises like the COVID-19 pandemic. The pandemic especially highlighted the vulnerability of the educational system worldwide, where global lockdowns and restrictions disrupted access to laboratory facilities, severely impacting science-based laboratory education ~\cite{unesco2019, worldbank2020}. These constraints highlight the urgent need for innovative, flexible solutions to ensure uninterrupted, quality education under diverse circumstances. Remote lab is one such practical solution, which is the focus of this paper. 

\subsection{Overview of Remote Labs}
\begin{figure}[tb]
    \centering
    \includegraphics[width=0.6\columnwidth]{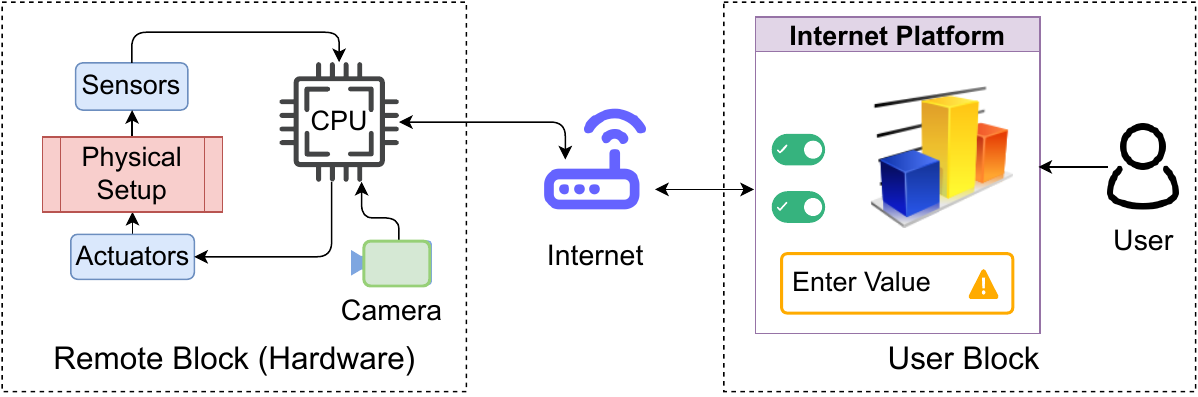}
    \caption{Overview of an IoT-based remote lab}
    \label{fig:flow}
    \vspace{-4mm}
\end{figure}

In remote labs, laboratory experiments situated at a given geographical location can be accessed from anywhere and anytime in the world using the internet over a browser. Fig. \ref{fig:flow} shows the overview of a generic IoT-based remote lab, which consists of two blocks connected by the internet. The first block consists of the internet platform at the user's side. 
The second block consists of hardware at the remote location and includes the physical setup of the experiment, sensors, actuators, cameras, and controllers. In such setups, cameras are used to stream the live feed of the experiment to the user. The user can control the experiment parameters using the internet platform. Experiment results can be presented in various formats, including tables, interactive plots, and diagrams on the internet platform.


\subsection{Related Work}

Numerous universities and research groups worldwide have implemented IoT-based remote lab solutions \cite{labsland_adapt,labsland_replication,labsland_wilsp,ises,labsland_lowcost,laborem,blockino,RaspyLab,RaspyControl,uned_control,edinburgh,netherlands_control}. Prominent examples include WebLab-Deusto and its spin-off, LabsLand, both originating from the University of Deusto \cite{labsland_lowcost,labsland_reliable,labsland_wilsp}. Their design of remote labs in different science and engineering domains comes with extensive features that primarily focus on building low-cost and reliable experiments by creating multiple instances of an experiment. 
UNILabs\footnote{\url{https://unilabs.dia.uned.es/}} \cite{unedlabs, fislabs} is a network of web-based laboratories with several Spanish universities participating and hosting experiments that are shared over a Learning Management System \cite{automatlabs}. 

Additionally, the Internet School Experimental System (iSES) \footnote{\url{https://www.ises.info/index.php/en}} offers a wide range of experiments in physics and allied domains. The iSES Remote Lab SDK, comprising customisable JavaScript widgets and low-latency communication, enables programmers and non-programmers to create remote laboratory interfaces with data and video transfer facilities \cite{ises}. 

Specialised iterations of remote labs have emerged to cater to specific domains of science and engineering, including electronics and embedded systems \cite{labsland_lowcost, laborem}. In \cite{labsland_lowcost}, a multi-instance microcontroller-based embedded device remote laboratory is developed, surpassing state-of-the-art architectures in terms of scalability on low-cost setups. Furthermore, \cite{laborem} introduces an open-source interface equipped with interchangeable circuit boards (plugs), providing educators with the flexibility to adapt circuits swiftly. Remote labs also teach programming languages like Python \cite{blockino, RaspyLab}, enabling users to program sensors and actuators, such as temperature sensors and LCDs. 
Several implementations of remote labs in the control engineering domain are discussed in \cite{RaspyControl, uned_control, edinburgh, netherlands_control}. In \cite{RaspyControl}, the authors introduce a low-cost, fully open-source remote laboratory for teaching automatic control systems. It includes complete instructions for building and replicating the remote laboratory's hardware and software components. In \cite{uned_control}, a low-cost remote laboratory for control engineering experiments is developed using an inexpensive BeagleBone Black development board\footnote{\url{https://www.beagleboard.org/boards/beaglebone-black}}, with experiments hosted on a dedicated Java application. In \cite{edinburgh, netherlands_control}, authors discuss the use of remote laboratories in control engineering education experiments, highlighting the importance of active learning pedagogy while providing limited insight into the implementation details.

\subsection{Attributes of Remote Labs}

As mentioned previously, the literature presents numerous remote lab solutions (systems), each with its features and capabilities. To establish the suitability of these systems, the non-functional attributes (NFAs), also referred to as non-functional requirements \cite{nfa}, are considered. There are several NFAs, and they define how a system will work by laying out its requirements and limitations. It is crucial to establish a common ground, and to this end, this paper has identified seven crucial NFAs and classified relevant features under each NFA. These NFAs and features serve as critical comparison criteria for remote laboratory solutions. The identified NFAs and their features are as follows:

\begin{enumerate}

    \item \textbf{Affordability}: It describes the cost-effectiveness of a remote labs system, primarily focusing on \textit{low-cost designs}. The cost factor is particularly crucial for making labs affordable and accessible to a large number of students in developing countries like India.
    
    \item \textbf{Portability}: It describes the ease with which hardware experiments can be transported from one location to another. This can be achieved by designing \textit{modular setups}.
    
    \item \textbf{Scalability}: It describes the system's capacity to grow and adapt to increasing demands.
     This can be achieved in multiple ways:
    \begin{itemize}
        \item \textit{Adding hardware (HW) instances to the platform} - denotes the capability of adding multiple instances of the same experiment to the remote lab's platform.
        \item \textit{Miniaturised HW setups} - refers to designing remote experiments that are more compact than traditional laboratory experiments.
        \item \textit{Peer-to-Peer (P2P) live video streaming} - refers to a live video streaming architecture that does not require a centralised streaming server.
    \end{itemize}
    
    \item \textbf{Compatibility}: It describes the system's ability to integrate and operate with a diverse range of configurations. This includes several features of the system:
    \begin{itemize}
        \item Compatibility with \textit{different hardware boards} for controlling the actuators and sensors and connect to the internet for data transmission (like ESP32\footnote{\url{https://www.espressif.com/en/products/socs/esp32}} and Raspberry Pi\footnote{\url{https://www.raspberrypi.com/products/}}).
        \item Compatibility with \textit{multiple IoT platforms} for receiving data from different hardware boards and making the data accessible online (like Blynk\footnote{\url{https://blynk.io/}} and Thingspeak\footnote{\url{https://thingspeak.com/}}).
        \item Compatibility with \textit{various types of cameras} for live video streaming of the experiments. Cameras can include standalone cameras like IP cameras and non-standalone cameras like USB cameras.
    \end{itemize}
    
    \item \textbf{Maintainability}: It describes the efforts required to keep the system running normally. \textit{Automated testing} showcases the system's ability to automatically report any errors affecting its functionality.
    
    \item \textbf{Usability}: It describes how user-friendly the system is and is best known from \textit{user feedback} collected from individuals (users) who have used the system.

    \item \textbf{Universality}: It refers to the system's ability to be accessed across a variety of devices (mobiles, desktops), operating systems (Windows, iOS, Android), and web browsers (Firefox, Edge, Chrome), maintaining a consistent and adaptable user experience.

\end{enumerate}

These NFAs offer a comprehensive basis for evaluating and qualitatively comparing various remote lab solutions, providing valuable insights into different solutions' strengths and weaknesses. Recent works ($\leq$ eight years) that provide proper implementation details of the hardware and software components have been considered to evaluate prominent and relevant remote lab solutions in the literature. The focus is on labs operating within web browsers without external plugins rather than dedicated applications. According to the above criteria, the chosen remote lab implementations are:
LabsLand \cite{labsland_lowcost,labsland_wilsp,labsland_reliable}, RaspyLab \cite{RaspyLab}, RaspyControl Lab \cite{RaspyControl} and iSES \cite{ises}. Table \ref{tab:lit_comparison} compares the chosen remote labs among the identified NFAs and the proposed solution in this paper.

\begin{table*}[tb]
    \caption{Feature comparison of various Remote Labs}
    \label{tab:lit_comparison}
\centering
\begin{tabular}{|c|c|c|c|c|c|c|}

\hline
\rotatebox{\headerangle}{\parbox{\headerlength}{\centering Attribute}} & \rotatebox{\headerangle}{\parbox{\headerlength}{\centering Features}} & \rotatebox{\headerangle}{\parbox{\headerlength}{\centering LabsLand\\\cite{labsland_lowcost,labsland_wilsp,labsland_reliable}}} & \rotatebox{\headerangle}{\parbox{\headerlength}{\centering RaspyLab\\\cite{RaspyLab}}} & \rotatebox{\headerangle}{\parbox{\headerlength}{\centering RaspyControl\\\cite{RaspyControl}}} & \rotatebox{\headerangle}{\parbox{\headerlength}{\centering iSES\\\cite{ises}}} & \rotatebox{\headerangle}{\parbox{\headerlength}{\centering \textbf{RLabs}\\\text{(Proposed)} }} \\
\hline

\textbf{Affordability}                    & Low-cost setups                              & $\checkmark$      & $\checkmark$      & $\checkmark$              & $\times$   & $\checkmark$  \\ \hline
\textbf{Portability}                      & Modular hardware setups           & $\times$       & $\times$       & $\times$               & $\times$   & $\checkmark$  \\ \hline
\multirow3{*}{\begin{tabular}[c]{@{}c@{}}\textbf{Scalability}\end{tabular}} & Adding HW instances to the platform & $\checkmark$ & $\checkmark$ & $\times$ & $\checkmark$ & $\checkmark$ \\ {}
                                 & Miniaturised HW setups                       & $\times$       & $\times$       & $\times$               & $\times$   & $\checkmark$  \\ {}
                                 & Peer-to-Peer live video streaming                    & $\times$      & $\times$      & $\times$               & $\times$  & $\checkmark$   \\ \hline
\multirow{3}{*}{\textbf{Compatibility}}   & Tested different HW boards                       & $\checkmark$      & $\times$       & $\times$               & $\checkmark$  & $\checkmark$  \\ {}
                                 & Tested multiple IoT platforms               & $\times$       & $\times$       & $\times$               & $\times$   & $\checkmark$  \\ {}
                                 & Tested different cameras                          & $\checkmark$      & $\times$       & $\times$               & $\times$   & $\checkmark$  \\ \hline
\textbf{Maintainability}                   & Automated testing                     & $\checkmark$       & $\times$       & $\times$               & $\times$   & $\checkmark$  \\ \hline
\textbf{Usability}                        & User feedback                         & $\checkmark$      & $\checkmark$      & $\times$               & $\checkmark$  & $\checkmark$  \\ \hline
\textbf{Universality}                     & Device-friendly                       & $\checkmark$      & $\checkmark$      & $\checkmark$              & $\checkmark$  & $\checkmark$  \\ \hline

\end{tabular}%
\text{}
\\[1\baselineskip]
\checkmark  Implemented  \qquad $\times$  Not Implemented 
\end{table*}

\subsection{Contributions of the Paper}

 The contributions of this paper are as follows:

\begin{itemize}

    \item An end-to-end remote labs solution, known as \textbf{RLabs} (short for \textbf{R}emote \textbf{Labs}), has been implemented and deployed on the campus of the International Institute of Information Technology (IIIT) in Hyderabad, India \footnote{\url{https://www.iiit.ac.in/}}.
    \item Two use case experiments, Vanishing Rods and Focal Length, based on high-school physics, are designed by retrofitting IoT components to the traditional laboratory equipment.

 \renewcommand{\thefootnote}{\fnsymbol{footnote}} 
    \item Additionally, modular and miniaturised versions of the two use case experiments are also proposed and built{\footnote[1]{Initial results are presented in a conference publication \cite{das2023}}} to make the setups portable and scalable.

    \item A novel software architecture for RLabs has been proposed and implemented. This architecture offers several novel features that facilitate the seamless scalability of remote experiments. This includes the introduction of an interoperability layer to support multiple IoT platforms, and the implementation of P2P live video streaming.
    \item An automatic testing system is designed to verify if the remote experiments are functioning as intended, eliminating the need for manual checks. This system includes the usage of CV algorithms and Selenium-based automation and is integrated with the proposed RLabs software architecture.
    \item User feedback from forty-five students located at a different geographical city is collected and reviewed, containing questions to rate their usability and learning outcomes after using the built RLabs system.
    
\end{itemize}

\renewcommand{\thefootnote}{\arabic{footnote}}

\par

The proposed work is novel concerning the relevant recent remote lab implementations as shown in Table \ref{tab:lit_comparison}. The existing implementations lack features like modular and miniaturised hardware experimental setups, and P2P live video streaming and did not test different IoT platforms, leading to portability, scalability and compatibility limitations. Additionally, the systems in \cite{ises,RaspyControl,RaspyLab} have not tested different cameras for live-streaming and lack an automation testing suite that limits the compatibility and maintainability of the systems. In \cite{ises}, the setups are not low-cost as they use personal computers for connecting the experiments to the internet, affecting the affordability of the systems. In \cite{RaspyControl}, different hardware instances of an experiment cannot be added, and a user feedback is not presented that impacts the scalability and questions the system's usability. In \cite{RaspyLab}, different hardware boards are not tested to design the experiments that limit the system's scalability.

\subsection{Structure of the Paper}

The structure of the paper is outlined as follows. In Section \ref{sec:model}, the theory of the two use case experiments is briefly presented. Section \ref{sec:hardware} provides detailed insights into the hardware components used for retrofitting the experiments and creating their miniaturised versions. Section \ref{sec:software} elaborates on the software platform designed for hosting and managing these experiments. In Section \ref{sec:results}, results are articulated systematically by analysing the performance and effectiveness of our model against each proposed attribute. Finally, Section \ref{sec:conc} concludes the work presented.

%% file: source_model.tex
\par In this section, two considered use case experiments based on fundamental high-school physics concepts are explained in detailed, including their aim, theory, methodology, and results.

\subsection{Experiment 1: The Vanishing Rod Experiment}

\subsubsection{Aim:}
\textit{To observe the change in visibility of a glass rod when immersed in oil and in water media}

\begin{figure}[tb]
\begin{subfigure}[b]{0.565\columnwidth}
\centering
\includegraphics[height=3.7cm,width=\columnwidth]{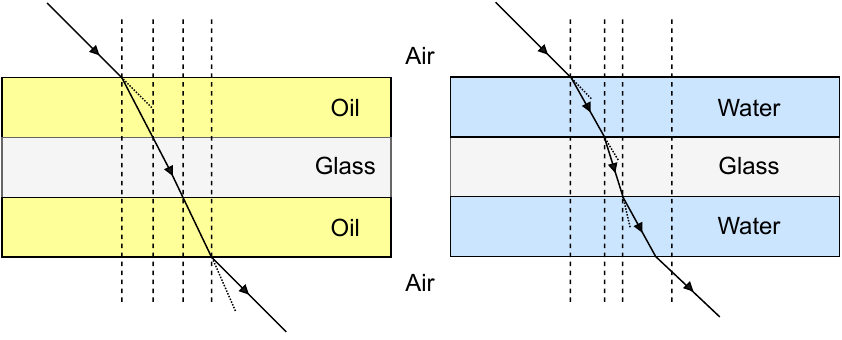}
\caption{}
\label{fig:vr_ray}
\end{subfigure}
\hfill
\begin{subfigure}[b]{0.4\columnwidth}
\centering
\includegraphics[height=3.6cm,width=\columnwidth]{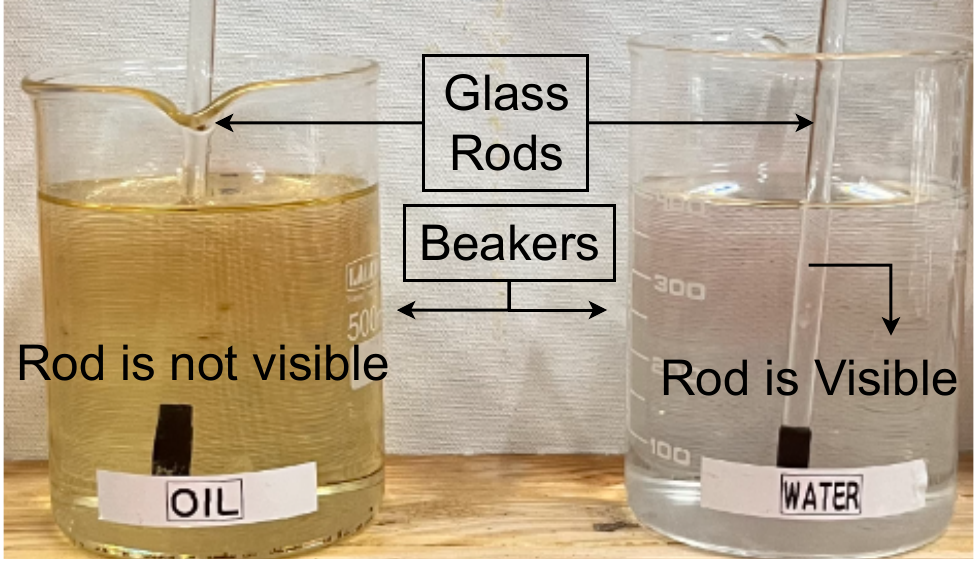}
\caption{}
\label{fig:vr_vanish}
\end{subfigure}
\hfill
\caption{(a) Ray diagram of light entering different surfaces. (b) Visibility of glass rods when dipped in beakers containing water and oil separately}
\label{fig:vr_cam}
\vspace{-4mm}
\end{figure}

\subsubsection{Theory:}
The concept of refractive index is explored in this experiment. While light travels from one medium (e.g. air) to another (e.g. water), the bending of light can be observed. This happens due to the slowing down of the speed of light when the medium changes. Every material is associated with a refractive index that quantifies the refraction of light. The higher the refractive index, the higher the deviation of light in the entering medium. However, suppose another medium with a numerically closer refractive index surrounds an object. Then, the object appears to have disappeared in the medium, as there will be no reflection and refraction of the light while passing through the object, as illustrated in Fig. \ref{fig:vr_ray}.

\subsubsection{Methodology:}
Fig. \ref{fig:vr_vanish} shows the apparatus used in the setup which includes borosilicate glass rods (Refractive index $\mu$ = 1.5), sunflower oil ($\mu$ = 1.47), drinking water ($\mu$ = 1.36), borosilicate glass beakers ($\mu$ = 1.5). One beaker is filled with sunflower oil and the other beaker is filled with water. The experiment is performed by placing the glass rods into the beakers and observing the visibility of glass rods in the beakers.

\subsubsection{Results:}
Fig. \ref{fig:vr_vanish} depicts the visibility of glass rods when immersed in glass beakers. It can be observed that the glass rod dipped in the oil beaker tends to disappear while the other glass rod remains clearly visible. This is attributed to the fact that the glass rod and sunflower oil have very similar refractive index values that make the glass rod vanish in the oil medium. However, the glass rod and water have different refractive index values that make the glass rod visible in the water medium.

\begin{figure}[tb]
\begin{subfigure}[b]{0.54\columnwidth}
\centering
\includegraphics[height=4cm,width=\columnwidth]{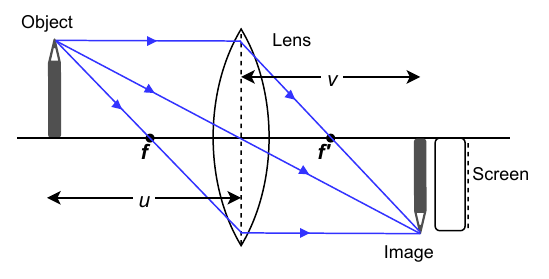}
\caption{}
\label{fig:ray}
\end{subfigure}
\hfill
\begin{subfigure}[b]{0.44\columnwidth}
\centering
\includegraphics[height=4cm,width=\columnwidth]{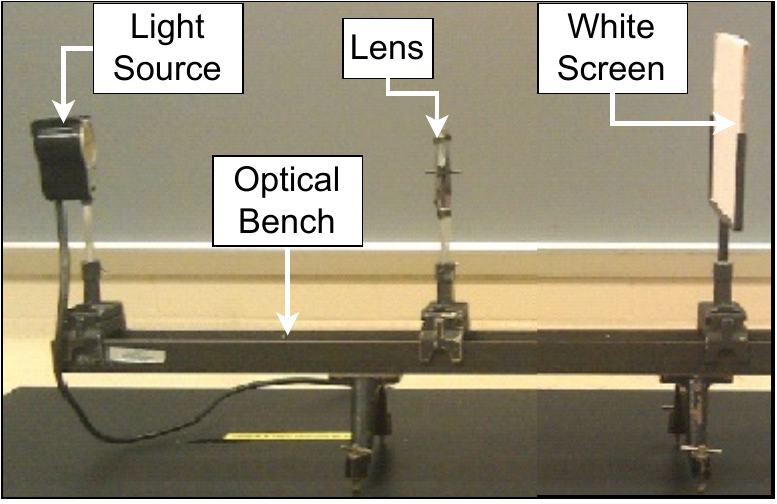}
\caption{}
\label{fig:fl_img}
\end{subfigure}
\hfill
\caption{(a) Ray diagram of a thin biconvex lens. (b) Experimental setup of the Focal Length experiment (source: \url{hirophysics.com})}
\label{fig:fl_exp}
\vspace{-4mm}
\end{figure}

\subsection{Experiment 2: Focal Length Experiment}

\subsubsection{Aim:}
\textit{To determine the focal length of a biconvex lens by forming a sharp image of an object on the screen.}  

\subsubsection{Theory:}
The experiment focuses on the field of optics and aims to determine the focal length of a biconvex lens \cite{thinlens}. 
The focal length is a measure that defines the distance between the lens and the point where light rays converge to form a sharp image. It determines the lens's viewing angle and the magnification of the image produced. 
To determine the focal length of a given thin biconvex lens, an experiment is set to get a sharp image of an object on a screen by adjusting the distances of the screen and object from the pole of the lens as shown in Fig. \ref{fig:ray}. Once a sharp image is formed on the screen, the object distance (\textit{u}) and the image distance (\textit{v}) are used to calculate the focal length (\textit{f}) of the used lens using the lens formula with proper sign conventions. 
Any measurement towards the direction of the incident ray is considered positive, and distances opposite to the direction of the incident ray are negative. In the case of a thin biconvex lens, as shown in Fig. \ref{fig:ray}, the object is on the left side of the lens and the real image will form on the right side if the object distance is greater than \textit{f}. So according to the sign convention, \textit{u} will always be negative and \textit{v} will always be positive for real images. Then the modified lens formula would be \(\frac{1}{f} = \frac{1}{v} + \frac{1}{u}\), where \textit{u,v} and \textit{f} would be the absolute values.

\subsubsection{Methodology}
Fig. \ref{fig:fl_img} shows the apparatus used in the setup which includes a white screen, a light source, a biconvex glass lens and an optical bench upon which the experiment is performed. Firstly, the biconvex glass lens is placed on the lens stand. The light source and the white screen are placed on the either sides of the lens such that the object, lens and the screen lie on a straight line. The experiment is performed by adjusting the positions of both the object and the screen platforms and observing the image formation on the screen.

\subsubsection{Results}

Table \ref{tab:fl_result_table} displays a few pairs of $u$ and $v$ values that yield a sharp image during the experimentation. It is notable that multiple sets of values can result in a clear and sharp image. Irrespective of these different values observed, the lens's focal length remains the same, given that the lens used is the same in every trial.

\begin{table}[tb]
\begin{minipage}{.49\textwidth}
  \centering
  \caption{Focal Length calculations for different values of \textit{u},\textit{v} for which sharp images are obtained}
  \begin{tabular}{|c|c|c|c|c|}
    \hline
    S.No & \textit{u} (cm) & \textit{v} (cm) & \textit{f} (cm) & \% error \\ \hline
    1 & 20.59 & 20.38 & 10.24 & 2.4 \\ \hline
    2 & 29.5 & 15.89 & 10.33 & 3.3 \\ \hline
    3 & 42.65 & 13.95 & 10.51 & 5.1 \\ \hline
  \end{tabular}
  \label{tab:fl_result_table}
\end{minipage}%
\hfill
\begin{minipage}{.49\textwidth}
  \centering
  \caption{Comparison of different stepper motors used in the experiments}
    \begin{tabular}{|l|l|l|}
      \hline
      \textbf{Motor Type} & \textbf{28BYJ-48} & \textbf{Nema 17} \\ \hline
      Step Angle & \ang{0.0875} & \ang{1.8} \\ \hline
      Rated Voltage (V) & 5 & 12-48 \\ \hline
      Rated Current (A) & 0.4 & 1.68 \\ \hline
      Holding Torque (kg-cm) & 0.34 & 4.2 \\ \hline
      Step Accuracy & Low & High \\ \hline
      Size (mm$^3$) & 34$\times$18$\times$10 & 40$\times$42$\times$42 \\ \hline
      Weight (g) & $\sim$50 & $\sim$300 \\ \hline
    \end{tabular}%
  \label{tab:motor_desc}
\end{minipage}
\end{table}

%% file: source_hardware.tex
In this section, hardware designs of retrofitted lab-scale remote experiments are presented along with their design choices. Later, more compact lab-scale versions are also presented, emphasising their modular and miniature designs.

\subsection{Retrofitted Lab-Scale Experimental Setups}

The lab-scale setups are created by retrofitting IoT components into the existing experimental equipment used in traditional laboratories, offering users a visual experience similar to traditional lab experiments.
\subsubsection{Experiment 1 - Vanishing Rod:}

\begin{figure}[tb]
\begin{subfigure}[]{0.29\columnwidth}
\centering
\includegraphics[height=4.3cm, width=\columnwidth]{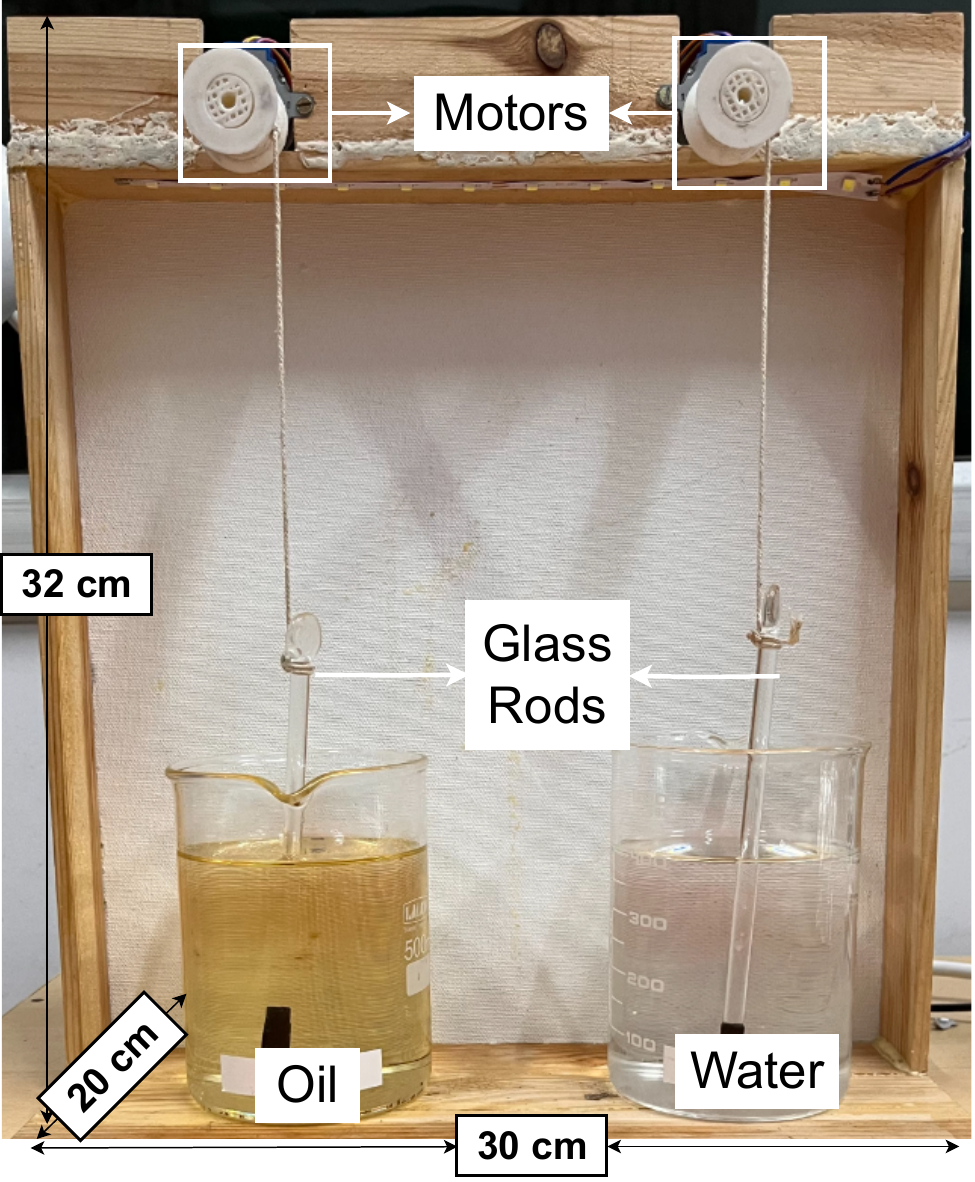}
    \caption{}
\label{fig:vr_setup}
\end{subfigure}
\hfill
\begin{subfigure}[]{0.32\columnwidth}
\centering
\includegraphics[width=\columnwidth]{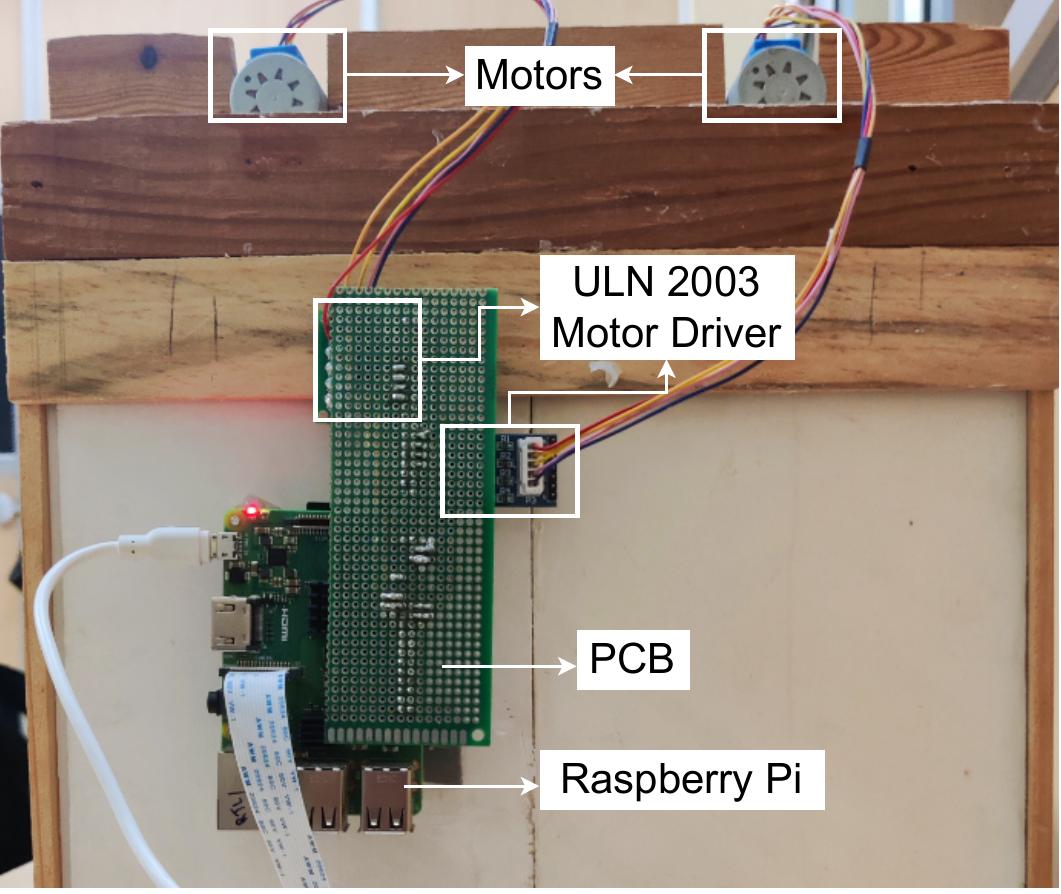}
    \caption{}
\label{fig:vr_back_setup}
\end{subfigure}
\begin{subfigure}[]{0.37\columnwidth}
\centering
\includegraphics[width=\columnwidth]{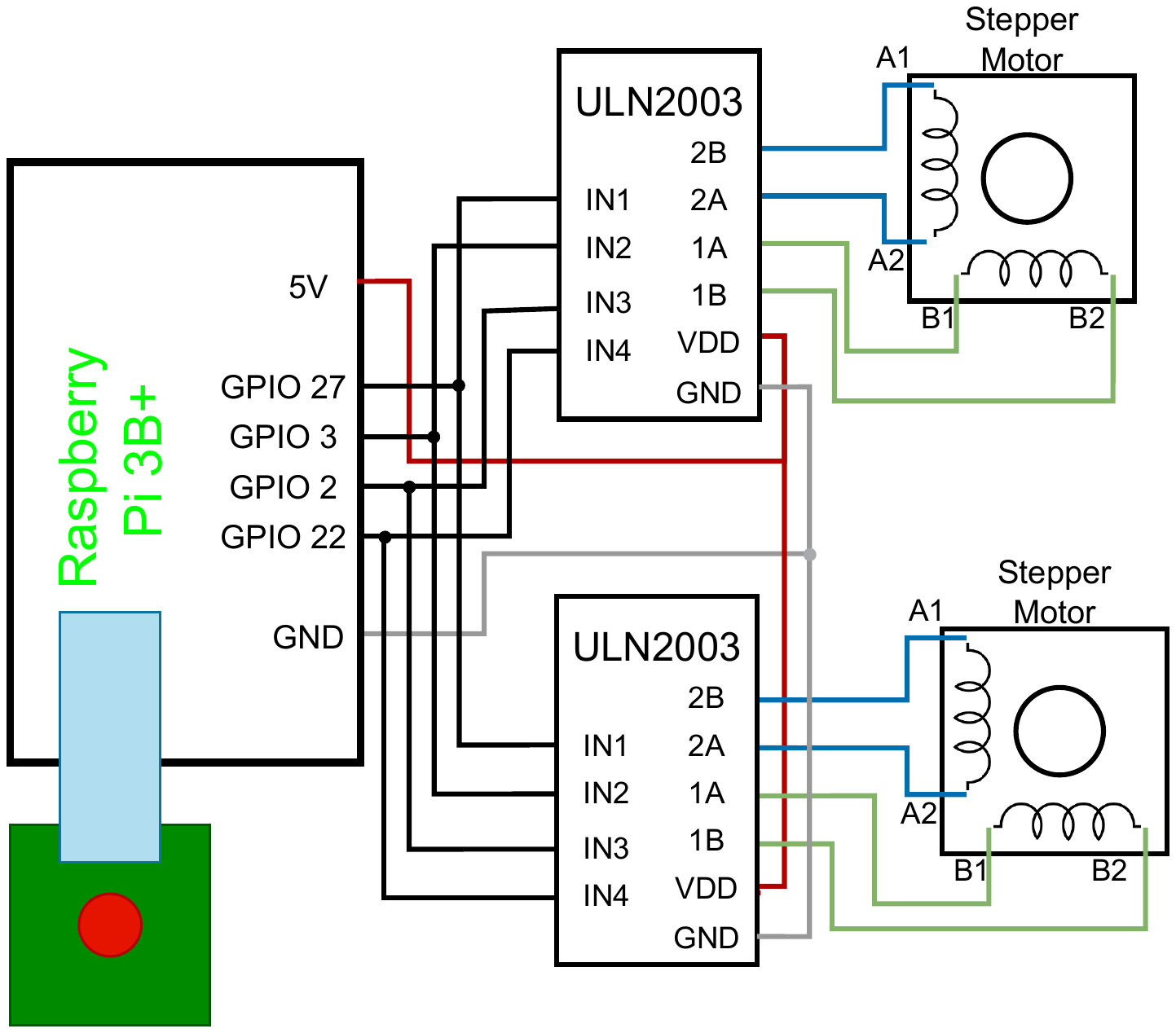}
    \caption{}
\label{fig:vr_ckt}
\end{subfigure}
\hfill
\caption{Hardware description of the lab-scale Vanishing Rod experiment. (a) Front-view of the experimental setup. (b) Back-view of the experimental setup. (c) Circuit diagram of the setup}
\label{fig:vr_hardware}
\vspace{-4mm}
\end{figure}

 The apparatus required for the traditional experiment is already discussed in Section \ref{sec:model}. Additional apparatus required to retrofit the experiment for remote accessibility includes plywood, 28BYJ-48 stepper motors (Table \ref{tab:motor_desc}), Raspberry Pi 3B+ and a Raspberry Pi camera (RaspiCam). Figs. \ref{fig:vr_setup}, \ref{fig:vr_back_setup}, and \ref{fig:vr_ckt} show the physical setup and circuit design of the Vanishing Rod experiment. One beaker is filled with sunflower oil, and the other is filled with water. They are then placed near the base, which is prepared from plywood. The glass rods are attached with strings to the stepper motors fixed above the beakers. The RaspiCam is positioned to capture both beakers within its field of view in the video feed. This feed is sent to Raspberry Pi 3B+, which controls the motors. The experiment is performed by moving the glass rods into the beakers placed and observing the visibility of the glass rods in the beakers. Both the glass rods are moved up or down vertically from the beakers simultaneously by the stepper motors connected to the Raspberry Pi. Users can control the glass rods' movement, and the glass rods can either be dipped into the beakers or not.

 \par The Raspberry Pi 3B+\footnote{\url{https://www.raspberrypi.com/products/raspberry-pi-3-model-b-plus/}} is chosen as the primary board for several reasons. Firstly, it boasts multiple GPIO pins, enabling interaction with numerous sensors and actuators simultaneously. Secondly, its compatibility with a Linux operating system facilitates programming in various languages. Thirdly, it supports various cameras like the RaspiCam and the USB cameras. Specifically, the RaspiCam, which is used for live-streaming the experiment, can be connected via the Camera Serial Interface (CSI) to the Raspberry Pi. The CSI is advantageous due to its high data throughput, ensuring real-time image and video processing \cite{csi}. Lastly, the Raspberry Pi 3B+ provides versatile internet connectivity options via Wi-Fi or an Ethernet port. Different types of motors can be used for the actuation, such as servo motors and stepper motors \cite{motors_comp}. Each motor serves a specific purpose, varying in accuracy, cost, and delivered torque. In this work, stepper motors are primarily utilised due to their cost-effectiveness, aligning with the requirements of the presented experiments. 

\subsubsection{Experiment 2 - Focal Length:}

\begin{figure*}[tb]
\includegraphics[width=0.85\columnwidth]{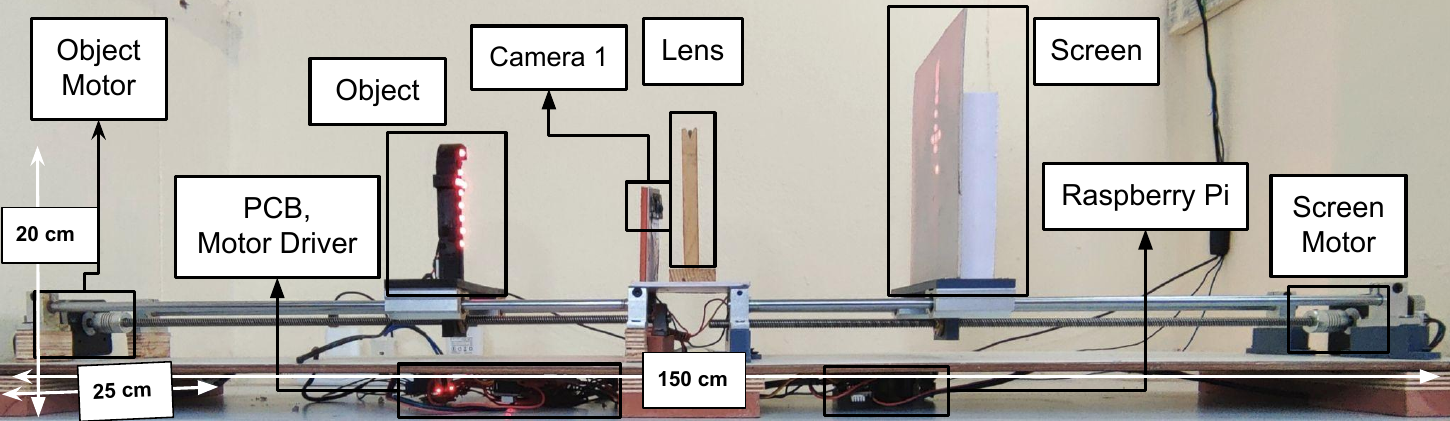}
\caption{Side-view of the experimental setup for Focal Length experiment}
    \label{fig:fl_side}
\end{figure*}

\begin{figure}[tb]
\includegraphics[width=0.45\columnwidth]{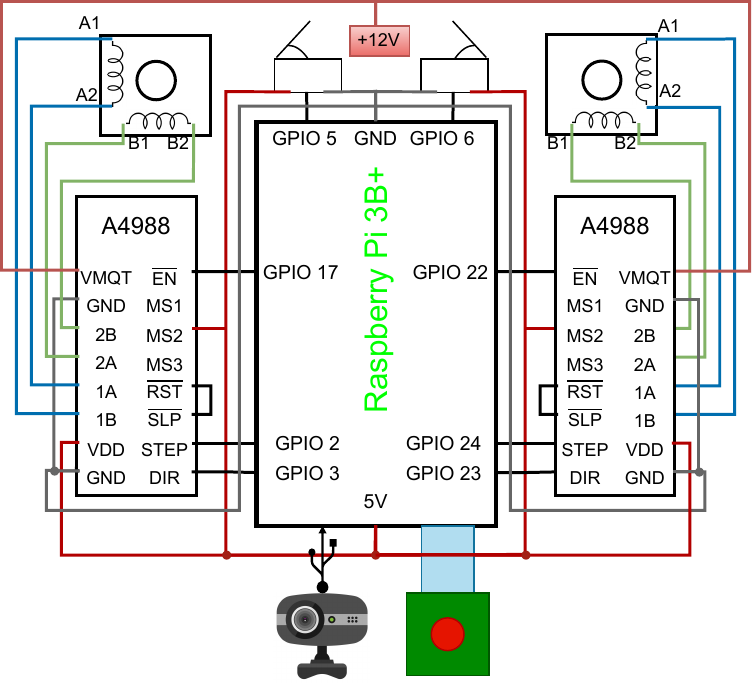}
\caption{Circuit diagram of the Focal Length setup}
    \label{fig:fl_ckt}
\end{figure}

The apparatus required for the traditional experiment is already discussed in Section \ref{sec:model}. Additional apparatus required to retrofit the experiment for remote accessibility includes two NEMA 17 stepper motors (Table \ref{tab:motor_desc}), two A4988 micro-stepping drivers, two limit switches (end-stop switches), a light source, a Raspberry Pi 3B+, a RaspiCam and an Ant Esports USB camera\footnote{\url{https://antesports.com/product/ant-esports-streamcam120-1080p-hd-webcam/}}. Figs. \ref{fig:fl_side} and \ref{fig:fl_ckt} show the experiment's physical setup and circuit design. Stepper motors are used to move the object and screen platforms horizontally while the position of the lens is fixed. The lens's focal length is calculated by considering the distances moved by the screen and object platforms from the lens once a sharp image is formed. 

\par The stepper motors used for this experiment differ from those used in the Vanishing Rod experiment as more accuracy and torque are required (Table \ref{tab:motor_desc}). The motors are attached to a screw shaft on which the screen/object is mounted, which converts the rotational motion to linear motion. Individual sliders facilitate the movement of the object and screen independently. The distance moved by the object and screen is linearly proportional to the number of steps/ degrees rotated by the stepper motor, which moves in precise and repeatable increments, which allows for consistent and accurate movement of the object and the screen. Micro-stepping drivers (A4988) control the motors, which improve the movement of the screen and object platforms. This micro-stepping driver allows the motor to take 400 steps per revolution. Here, 1 step is equivalent to \SI{10}{\micro\metre} and all the movements are based on this relation. 

\par  Limit switches are placed to recalibrate the motors that control the object and screen movements. The light source, which is electrically powered, is used as an object to observe the inversion in the image formed by the lens. The full-sized optical bench allows users to explore different image formations through the biconvex lens. RaspiCam is positioned to capture the white screen on which the image is formed, while the USB camera captures the side-view of the entire experimental setup. A USB camera is used instead of another RaspiCam as a single Raspberry Pi 3B+ can only handle one RaspiCam interfaced using CSI. The captured feed from both cameras is sent to the Raspberry Pi, which also controls the motors and receives signals from the limit switches.


\subsection{Miniaturised Experimental Setups}

 Miniaturised setups are smaller replicas of the lab-scale setups, primarily constructed using 3D printed components and commonly available materials that require no welding, adhesive bonding, or specialised mechanical tools typically used to build lab-scale setups. The designs are intentionally crafted for easy assembly, allowing users to follow a manual for straightforward construction. Minimal tools, such as a screwdriver and soldering iron, are sufficient for assembly.

\subsubsection{Experiment 1: Vanishing Rod:} 

\begin{figure}[tb]
    \centering
    \begin{subfigure}[b]{0.33\textwidth}
        \centering
        \includegraphics[height=5.5cm,width=\textwidth]{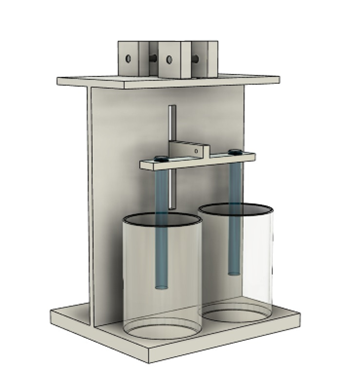}
        \caption{}
        \label{fig:vr_mini}
    \end{subfigure}
\hfill
    \begin{subfigure}[b]{0.3\textwidth}
        \centering
        \includegraphics[height=5.5cm,width=\textwidth]{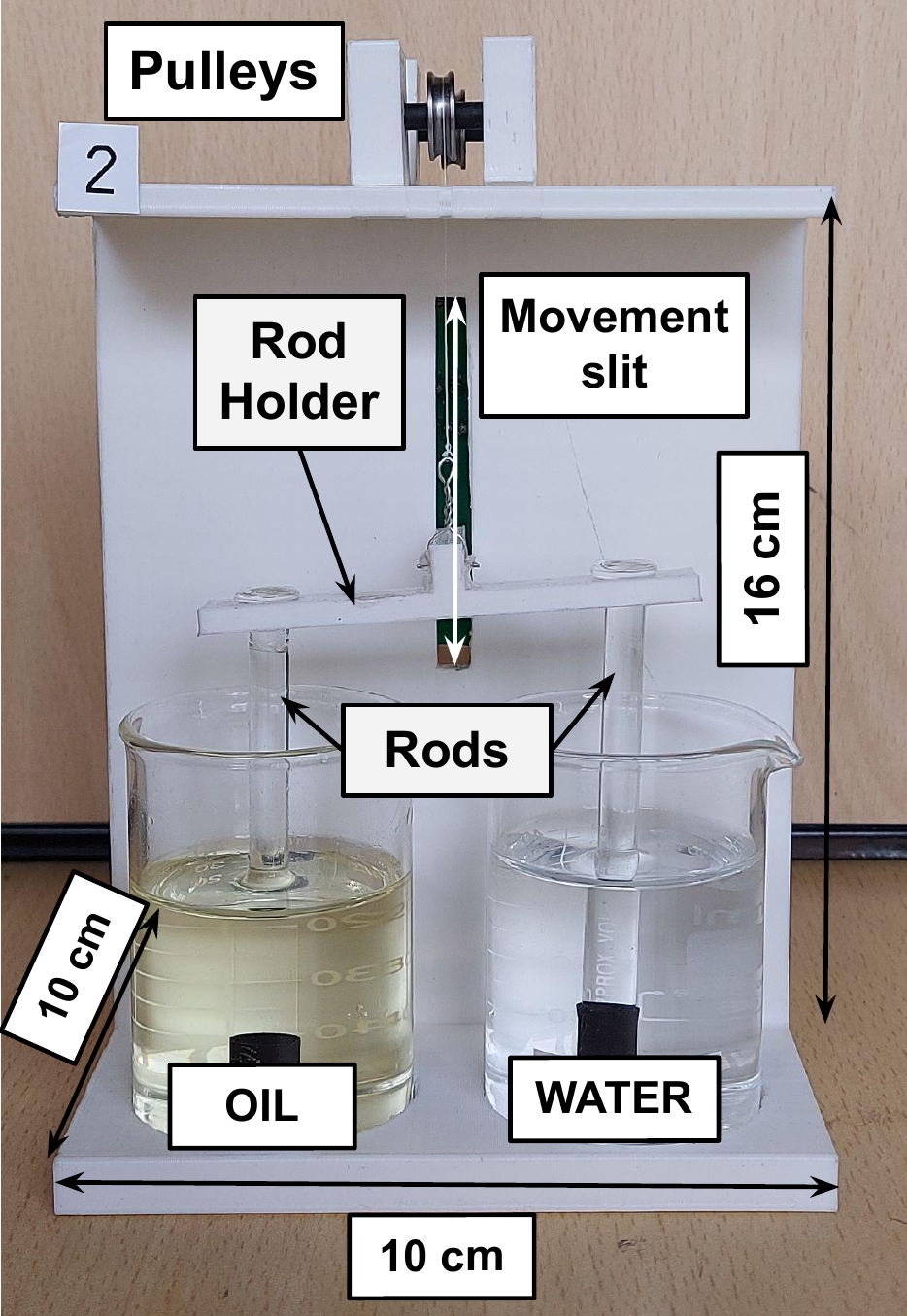}
        \caption{}
        \label{fig:minivr_front_view}
    \end{subfigure}
\hfill
    \begin{subfigure}[b]{0.3\textwidth}
        \centering
        \includegraphics[height=5.5cm,width=\textwidth]{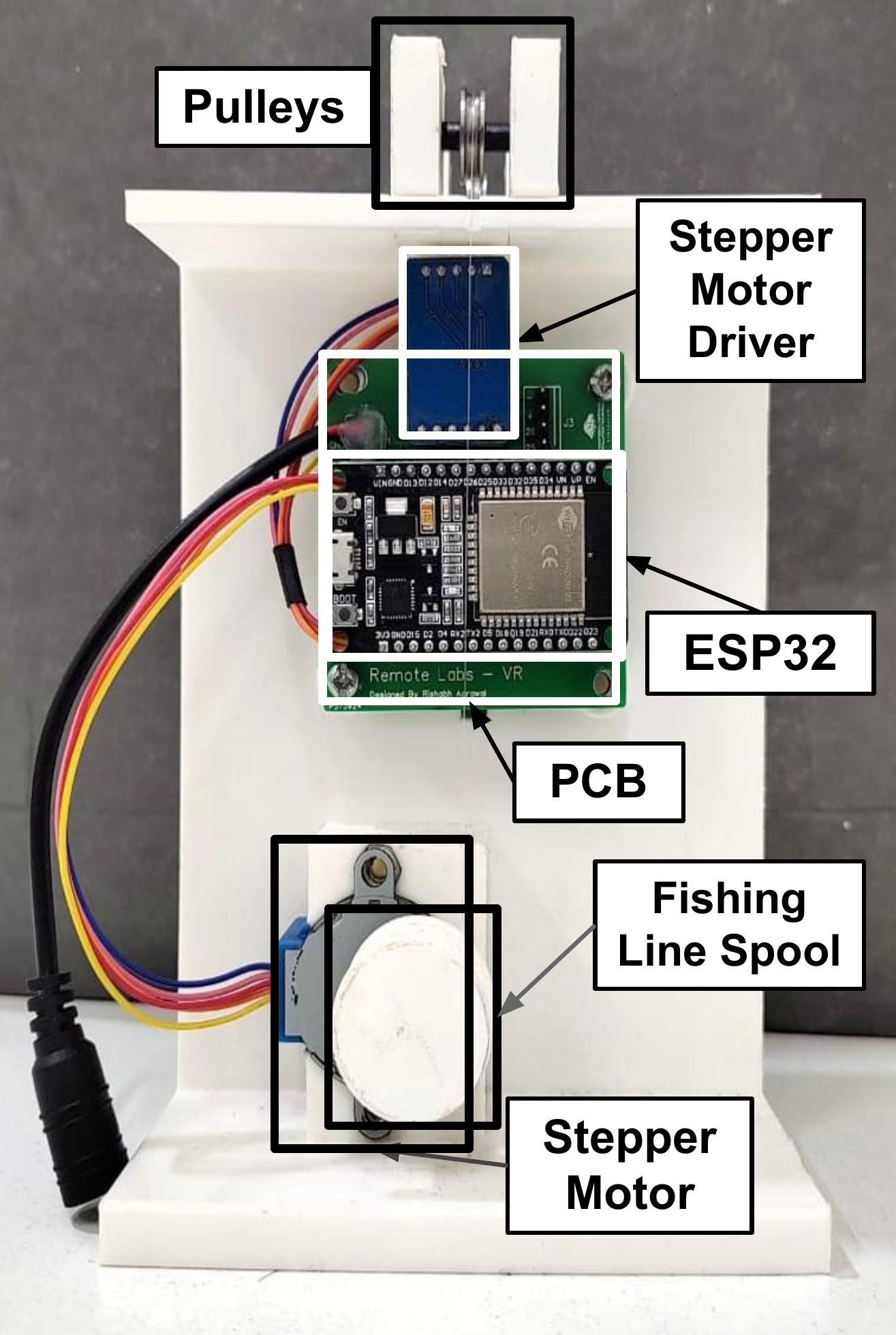}
        \caption{}
        \label{fig:minivr_back_view}
    \end{subfigure}
    \caption{Hardware description of the Miniaturised Vanishing Rod setup. (a) 3D model of the setup. (b) Front-view of the setup. (c) Back-view of the setup}
    \label{fig:vr_miniature_view}
\end{figure}

Fig. \ref{fig:vr_miniature_view} shows the miniaturised Vanishing Rod experiment. The miniaturised setup primarily consists of an ESP32, Raspberry Pi Zero 2 W, a single 28BYJ-48 stepper motor, ULN2003 motor driver, two pulleys, two borosilicate glass rods, and two 50mL beakers filled with sunflower oil and water. The exoskeleton of the experiment is entirely 3D printed in multiple smaller parts that can be assembled like puzzle pieces within a few minutes. ESP32 board controls the hardware components that save space without affecting any functionality of the experiment. Slots are provided on the exoskeleton to fit the beakers, glass rods and electronic components that are part of a single PCB, as shown in Fig. \ref{fig:minivr_back_view}. The glass rods are tied to a small rectangular platform (rod holder) spooled over a pulley connected to a single motor that rotates clockwise and anticlockwise. This platform is designed to move in a linear path — up and down. It is constrained into a slit which blocks it from rotating and toppling, keeping rods stable when moving, whereas, in the lab-scale experimental setup, the rods can freely rotate and swing when rods are moving. The slit which constrains this platform is indicated by the arrow in Fig. \ref{fig:minivr_front_view}. 
\par Raspberry Pi Zero 2 W is a more compact version of the Raspberry Pi 3B+ with a CSI port for video streaming and Wi-Fi connectivity. However, it has slightly lower processing capabilities and lacks an Ethernet port. It is important to note that the streaming pipeline is decoupled from the experiment controls in specific setups to showcase the compatibility of various hardware boards and cameras. This is discussed in detail in Section \ref{sec:results}.

\begin{figure*}[tb]
    \centering
    \includegraphics[width=\textwidth]{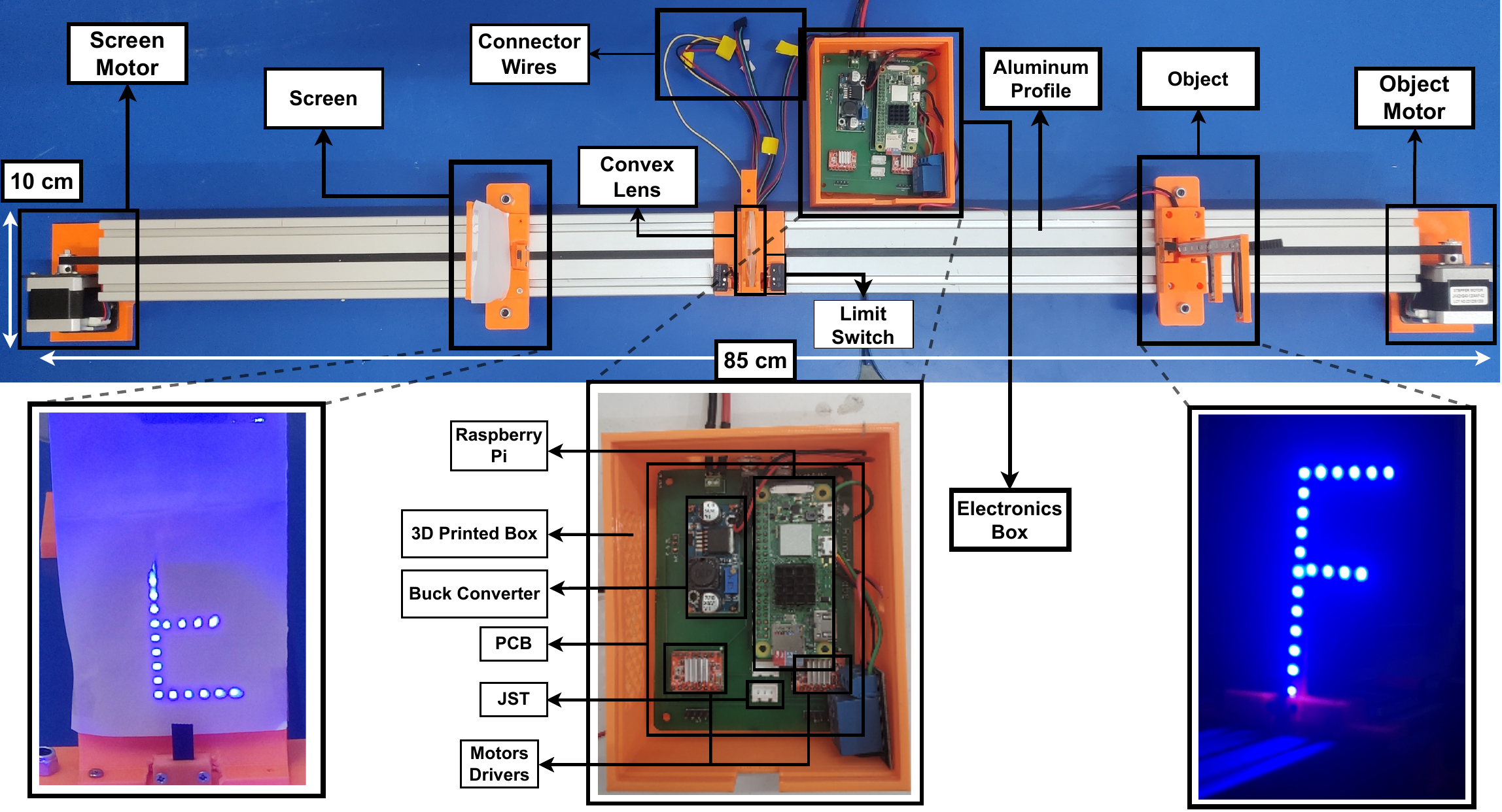}
    \caption{Hardware description of the Miniaturised Focal Length setup}
    \label{fig:fl_miniature_view}
    \vspace{-4mm}
\end{figure*}

\subsubsection{Experiment 2: Focal Length:} 
Fig. \ref{fig:fl_miniature_view} shows the miniaturised Focal Length experiment. The miniaturised setup apparatus employs a design similar to that of 3D printers, comprising two A4988 drivers, NEMA 17 stepper motors, limit switches, two Raspberry Pi Zero 2 W\footnote{\url{https://www.raspberrypi.com/products/raspberry-pi-zero-2-w/}} and two RaspiCams. The body is constructed from V-slot aluminium profiles, which serve as a versatile base for attaching various components. Two such profiles are used, one each for the object and screen. Using these aluminium profiles as a base, the rest of the structure is built, which includes the mechanism for moving the platform upon which the object/ screen is placed, space for motor operation and slots for the electronics. The other structural components, like the screen and object platforms and supports for the motors, are 3D-printed. These 3D-printed parts are designed to have slots and holes for screws appropriately, which can be fixed to the aluminium extrusion profiles using sliding nuts (slide and lock mechanism), eliminating the need for drilling holes in the profiles. In contrast to the traditional linear screw actuation, this setup utilises a belt drive mechanism to manoeuvre the object and screen platforms. This modification reduces the number of components and decreases the overall weight of the apparatus. A timing belt facilitates the movement of the screen, and object looped over a pulley and the stepper motor. Once the two profiles are built with the motors joined at the ends, the profiles are joined with a joint 3D-printed base upon which the lens stand is mounted. This 3D-printed base has connections between the motors and the illuminated object. These connections are made using JST and DuPont connector wires, which enable plug-and-play links to the electronics box containing the experiment's circuitry, as depicted in Fig. \ref{fig:fl_miniature_view}. The PCB inside the electronics box provides slots for connecting the Raspberry Pi, buck converter, and motor drivers. Additionally, a power supply not included in the setup can be directly connected through the power socket on the box's side. A Raspberry Pi and a RaspiCam are placed to stream the side-view of the experiment. A single Raspberry Pi could not handle the streaming from both a RaspiCam and a USB camera; hence, two Raspberry Pis were used.

%% file: source_software.tex
\begin{figure*}[tb]
    \centering
    \includegraphics[width=0.65\textwidth]{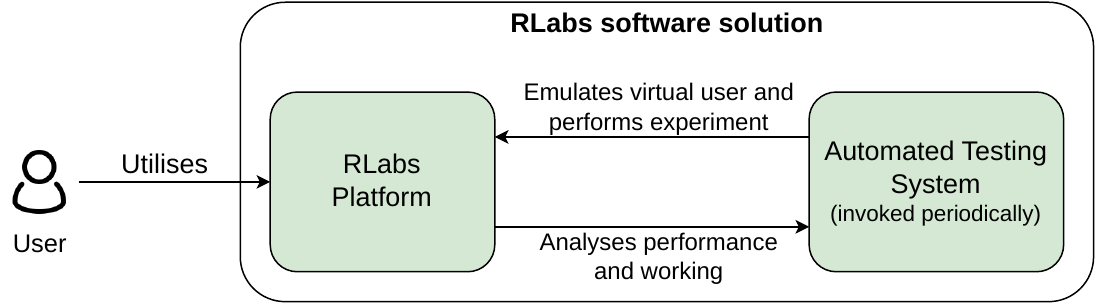}
    \caption{Overview of RLabs software solution}
    \label{fig:soln}
    \vspace{-4mm}
\end{figure*}

For the hardware devices described earlier, there is need for a software platform for user accessibility and a powerful and efficient dashboard for proper data management. 
As depicted in Fig. \ref{fig:flow}, the user interaction is enabled for such platforms via the internet platform. Remote lab solutions, including RLabs, are built on the foundational backbone that anyone can operate the hardware setups remotely. The internet platform developed as part of RLabs has been referred to has RLabs platform henceforth.
The RLab platform and devices must be accessible to anyone, anywhere and anytime \cite{ises}. For this purpose, the software platform must function appropriately on all smartphones, tablets, laptops and PCs without requiring much of additional applications \cite{hooke} for increased user accessibility. Moreover, for a platform enabling remote experimentation, ensuring the reliability of an experiment's outputs is crucial, as inconsistencies can impact the validity of the results obtained. This is particularly challenging in experiments that involve moving mechanical components due to the inherent complexities and potential for errors. In light of this, the RLabs platform is complemented by a comprehensive automated testing system, which has been developed to ensure the proper working of the platform as a whole. Fig. \ref{fig:soln} provides a high-level overview of the end-to-end RLabs software solution comprising the browser-based platform that facilitates the conduction of experiments remotely by the users, along with an automated testing system invoked periodically to emulate a virtual user and ensure that the platform works properly. A detailed description of the implementation is discussed in rest of the section.


\subsection{RLabs Platform}

\par
\begin{figure*}[tb]
    \centering
    \includegraphics[width=0.8\textwidth]{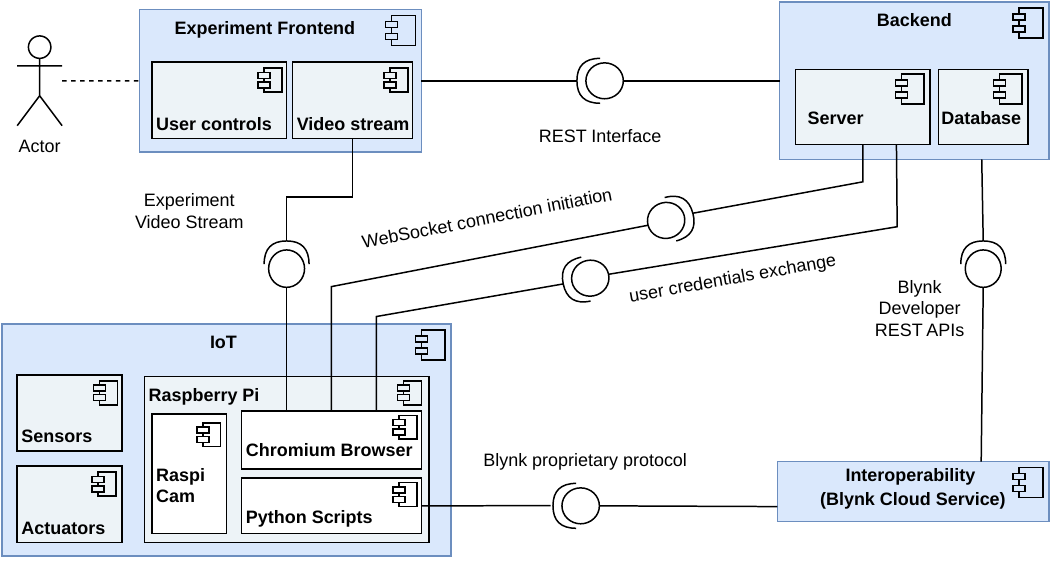}
    \caption{Major Components of RLabs Platform}
    \label{fig:components_sw}
    \vspace{-4mm}
\end{figure*}

 The proposed RLabs platform in Fig. \ref{fig:components_sw} presents four major components required for remote experimentation - \textit{the experiment frontend, the backend, the interoperability cloud service, along with the IoT component}. As the primary aim of RLabs is to conduct scientific experiments remotely based on user inputs, RLabs platform must have an interactive frontend for taking user inputs and a robust backend to relay these parameters and handle client requests. The communication between the frontend, backend and the experiments is a crucial aspect that must be considered while solutions requiring low-cost and low-latency are being built. An interoperability cloud service to facilitate the data exchange between multiple components and hardware devices using IoT for remote experimentation. Fig. \ref{fig:components_sw} uses standard UML (Unified Modeling Language) syntax to present an overview of these components and the interface between them. A standard way that the backend servers and software platforms, in general, employ to establish communication to the IoT component is by using HTTP(S) GET and POST requests to the interoperability layer. Moreover, HTTP(S)-based REST APIs provided by IoT platforms such as Blynk\footnote{\url{https://blynk.io/}} can be configured to work uniquely to our requirements and has been presented as the middleware and interoperability service in Fig. \ref{fig:components_sw}. It is important to mention that Blynk Cloud forming the interoperability component, can be substituted with any alternative if it provides the service of relaying the information between the backend and IoT component. The IoT component built for RLabs constitutes the  hardware like Raspberry Pi, sensors and actuators, which take the user inputs relayed by the interoperability component. Chromium browser in the Raspberry Pi is used to provide the WebSocket interface to the backend component.

\par RLabs software platform allows users to connect and control hardware remotely for conducting science experiments. Such platforms require the user to be able to view the outputs and results in real-time. This makes low-latency video streaming an essential feature of RLabs, especially when the platform involves the mechanical movement of hardware using actuators. As depicted in Fig. \ref{fig:components_sw}, the RLabs platform makes use of an open-source implementation of the WebRTC protocol to enable P2P communication and live stream the video feed of the RaspiCam captured by the Chromium browser directly to the user's frontend. Hence, user interaction with the experiments is possible in RLabs because of the perfect sync between the components and their integrated working creating a workflow unique to the framework, which has been explained in detail below.

\subsection{Execution flow}

\begin{figure*}[tb]
    \centering
    \includegraphics[width=\textwidth]{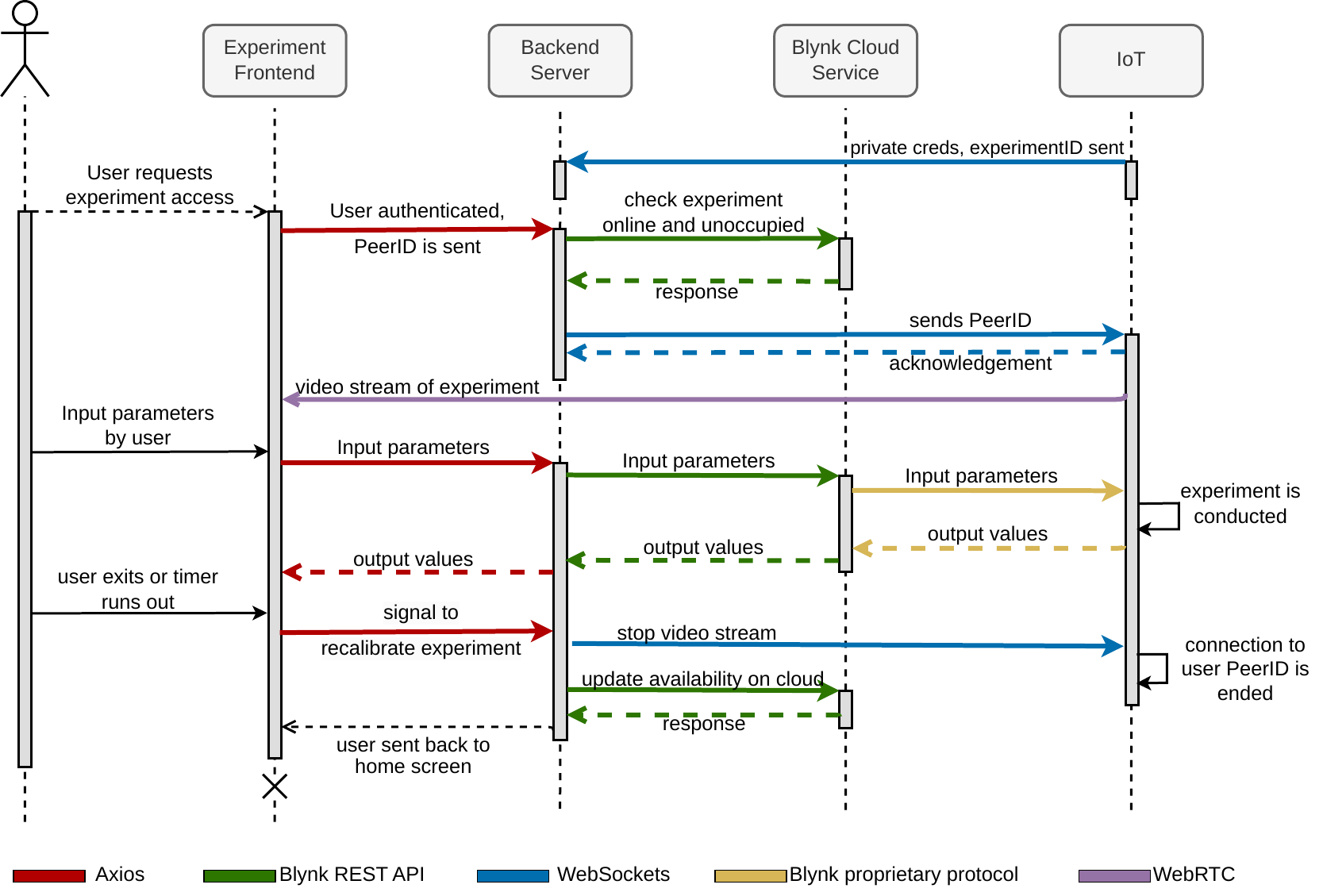}
    \caption{Sequence diagram for remote experimentation using the RLabs platform}
    \label{fig:seqeunce_sw}
    \vspace{-4mm}
\end{figure*}

Fig. \ref{fig:seqeunce_sw} shows the sequence diagram for the RLabs web-based platform following standard UML syntax with an appropriate legend to depict the communication protocol used between the components mentioned above for an unoccupied experiment hardware node. It provides a visual depiction of the execution flow and communication between the hardware and software components required to coordinate and relay information intricately to perform remote experimentation successfully using the holistic RLabs system. Any user experimentation comprises three initiation steps as depicted in Fig.\ref{fig:seqeunce_sw}- the user requesting experiment access, the user providing input parameters and the user exiting the experiment, each of which starts a unique communication flow. However, for this user-input and hardware-output-based system, the IoT embedded system first needs to inform the central backend server that it is active for remote experimentation which is performed by the Raspberry Pi of the experiment by sending its private credentials and experiment ID. An experiment ID is provided to each Raspberry Pi, to recognize which experiment a particular Raspberry Pi corresponds to. Moreover, some hardware nodes for experiments might have two or more Raspberry Pis sending the output value and video stream, due to which each Raspberry Pi requires its unique private credentials. By sending these two values, the Raspberry Pi is authorised by the central backend server, with the experiment hardware node now ready to take in user inputs, conduct experiments and generate output.

\par To perform an experiment, a user first must sign up and log in for verification and authorization. Once logged in, the user on the homepage can choose to access any experiment. As shown in Fig. \ref{fig:seqeunce_sw}, the first initiation step is where the user chooses the desired experiment to perform, after which the user is verified and authenticated again to prevent invalid access, and the PeerID generated automatically by the browser is sent to the backend server. Upon receiving the user PeerID successfully, a triple-fold check is performed for a hardware node's availability. A flag on the Blynk cloud is maintained for each experiment, which checks whether an experiment is under use. The connection of the hardware node with the Blynk platform for data exchange, flag value and the presence of a working RaspiCam stream are checked before marking an experiment as available and unoccupied for the current user. To check if the RaspiCam video stream is working, the user PeerID is sent to the Raspberry Pi, which responds with an acknowledgement if the service is active, and a WebRTC call with the video stream of the experiment is made by the Raspberry Pi directly to the user's frontend by connecting to the sent user PeerID. Once the experiment video stream from the Raspberry Pi has commenced, both frontend and IoT components are active and ready for user inputs. However, if an acknowledgement is not received in under five seconds, the experiment is marked unavailable and currently offline for experimentation.

\par The frontend client provides a user interface for users to interact with the hardware experiments. The second initiation step is when the frontend client provides inputs to a particular experiment using interactive buttons, sliders and switches, the server-client connection is established, and the information payload consisting of the input values is sent to the backend server. These are then communicated to Raspberry Pis associated with the hardware setups using the Blynk IoT platform. The experiment is conducted using the user-input parameters, and the output recorded is updated by the Raspberry Pis back on the Blynk platform. The server retrieves these outputs using GET API calls made to the Blynk Cloud, which are then communicated to the frontend, where they are displayed using graphs, charts, and tables as part of the graphical user interface (GUI).

\par Finally, when a user leaves, or the session times out, the experiment needs to recalibrate and be marked unoccupied so that it's available for the next user. For this purpose, a signal is sent to the Blynk Cloud to recalibrate the hardware experiment and change the flag value to unoccupied. The Raspberry Pi is also signalled to disconnect with the current user PeerID. The user is then redirected to the home page while the next client in the queue is permitted to enter this experiment and the flow repeats for each user remote experimentation. The components and features integral to the platform's implementation and enabling the student to use the interface are discussed in detail below.

\subsection{Key Architectural Design Decisions}

\begin{figure}[tb]
    \centering
    \includegraphics[width=0.8\columnwidth]{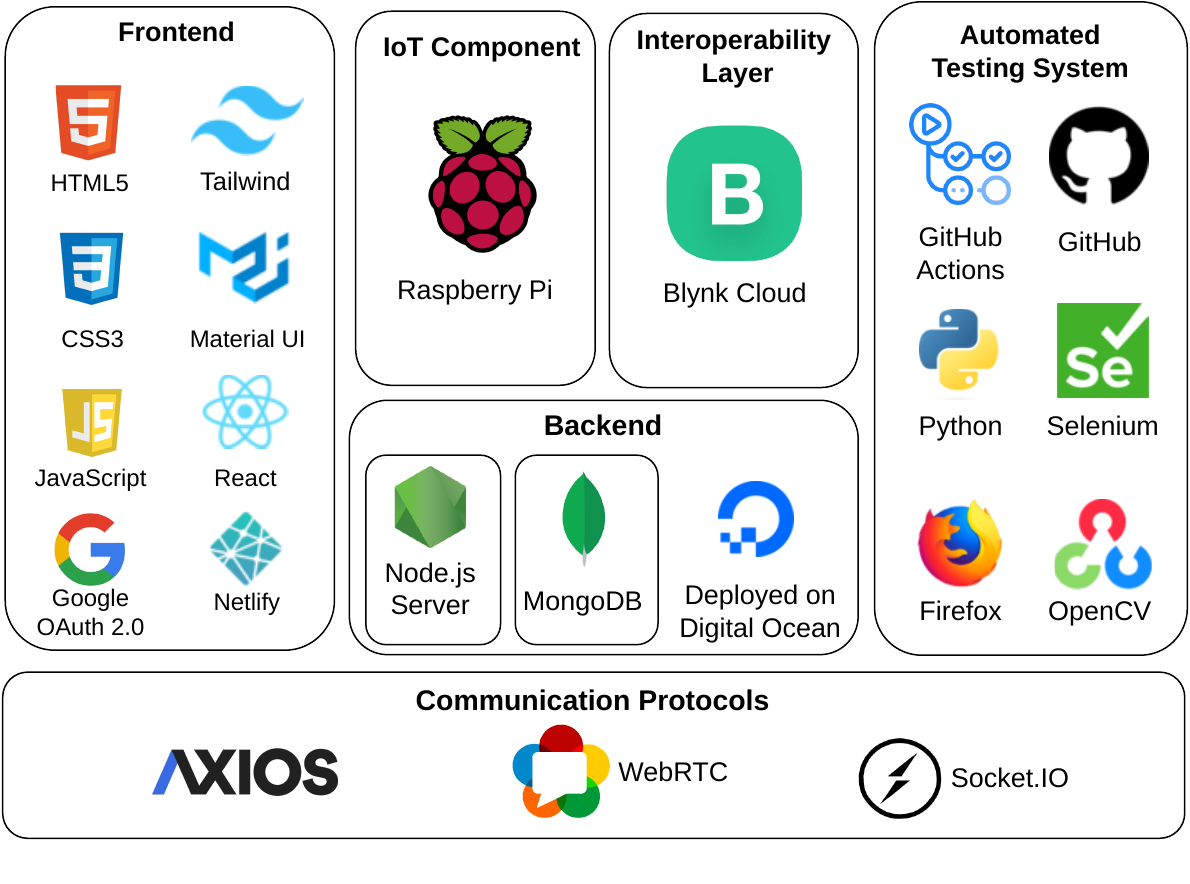}
        \vspace{-4mm}
    \caption{Technologies enabling RLabs platform}
    \label{fig:tech_stack}
    \vspace{-4mm}
\end{figure}

Fig. \ref{fig:tech_stack} presents the technologies used in implementing the four major components constituting RLabs, their subcomponents along with the protocols enabling communication between them and are explained in detail below.

\subsubsection{Experiment Frontend:}


The experiment frontend enables user interaction, allowing the client to provide input parameters through the buttons, sliders, and switches that form the user interface. Moreover, to view real-time outputs of the experiment, the users also require the experiment's video stream, which is crucial to any frontend enabling remote experimentation. The technologies used to implement the subcomponents - user interface and P2P video streaming in RLabs are provided in Fig. \ref{fig:tech_stack} and are explained below.

\paragraph {4.3.1.1 User Interface:}

The RLabs Document Object Model (DOM) and frontend are implemented using HTML5, CSS3 and JavaScript. HTML5 is a highly standardised language that helps provide structure to web pages and allows browsers to understand the content so that it can be displayed according to the styling and layout defined using the CSS3 language. Tailwind\footnote{\url{https://tailwindcss.com/}} and Material UI \footnote{\url{https://mui.com/material-ui/}} are styling libraries which speed up the development process by providing pre-built CSS classes for styling, offering the opportunity to implement customisable user interfaces. JavaScript allows for running event-driven logic in the student's browser \cite{ArPi}. React \footnote{\url{https://react.dev/}}, which is a popular component-based JavaScript library, allows us to develop and create advanced component-based user interfaces and is required as the platform has the potential to reuse UI components for different purposes. This allows for the frontend to be highly customisable and scalable. For the purpose of verification, both Google OAuth 2.0 has been integrated, and a local authentication system has been built. The Google OAuth 2.0 allows for seamless, smooth verification of the user. Axios \footnote{\url{https://axios-http.com/}}, a promise-based HTTP(S) client library, has been used to communicate between the frontend and backend as it provides the ability to intercept HTTP(S) requests from the browser or the server. As mentioned earlier, the platform's frontend is hosted on Netlify.


\paragraph{4.3.1.2 P2P Video Streaming:}

Live results are the foundation for conducting any physical experiment. The output changes must be visible when one tweaks any inputs during experimentation, enforcing minimal capture-render display in the video streaming service \cite{labsland_wilsp}. To live-stream the experiment and its outputs for added user experience, the platform uses WebRTC\footnote{\url{https://webrtc.org/}}. This modern open-source protocol allows real-time communication and data-sharing between two peers with minimal latency. Peers refer to the clients who exchange any information or data using WebRTC. Every client is identified using a unique PeerID. WebRTC, is a P2P protocol which means that it connects the video streaming sources with the consumer without any server joining them. It is noteworthy to mention that WebRTC does not use a central server for relaying the streaming data; however, it does use one for the signalling phase and connecting with the other peer. Moreover, after the signalling stage is completed and the peer is discovered, the data or stream happens entirely P2P. For the purpose of video streaming, an open-source implementation of the WebRTC protocol has been used, which uses the Node.js-based PeerJS server for the initial brokerage phase and the signalling to happen. The implementation captures the local stream of the experiments from the RaspiCam and the unique PeerID corresponding to the Raspberry Pi attached to the experiments. The P2P connection happens only when the user wishes to perform any experiment. WebRTC for our implementation has shown to have a latency of 200-400ms for one-way communication on a Raspberry Pi 3B+. 


\subsubsection{Backend Server:}

Using the MERN stack provides Express.js, a server-side web framework based on Node.js, which handles URL routing and asynchronous HTTP requests and responses. This helps connect with MongoDB, providing a scalable, document-based database to store all user credentials, experimental data, tokens, and environment variables. With a powerful Node.js-based backend server, the platform can be accessed easily once a user signs up, which can be done either locally or by connecting their Google accounts. All the authorization and verification details of the accounts signed up with the platform are stored securely in the MongoDB database. Only authorized clients are permitted to access the platform and use the hardware setups. Having a Node.js server makes our web application more user-friendly and accessible, as the features and implementation would depend on the server's node version, irrespective of the user's browser. The RLabs platform operates using a client-server connection established to access the experiments. The client-server communication utilises HTTP(S) with synchronous object transfer calls for transmitting and receiving data based on JSON using REST APIs. The asynchronous data flow is handled using "Asynchronous JavaScript and XML (AJAX)\cite{recon}," whereas the communication to the hardware setups takes place via Blynk. Hence, the server is tasked with handling communication with both the client and the Raspberry Pis, acting as an intermediate between them. As mentioned previously, the server and database has been hosted as a droplet on Digital Ocean for a paid annual subscription.

\subsubsection{Blynk Cloud:}

Forming the interoperability component, Blynk, the IoT platform provides services that are widely used to control hardware remotely, store and display remotely collected data. Blynk, through its developer mode\footnote{\url{https://docs.blynk.io/en/concepts/developer-mode}}, provides clients with the ability to modify and receive values using REST APIs from the cloud. It is this very feature which has been used extensively to communicate values to the hardware. When users try accessing and conducting a particular experiment, clients are connected to their desired experiment via Blynk. On providing the desired inputs using the platform, the values are modified on the corresponding Blynk device using REST APIs, which are then communicated to the Raspberry Pis of the corresponding experiments using Blynk's proprietary protocol.

\begin{figure}[tb]
    \centering
    \includegraphics[width=0.6\columnwidth]{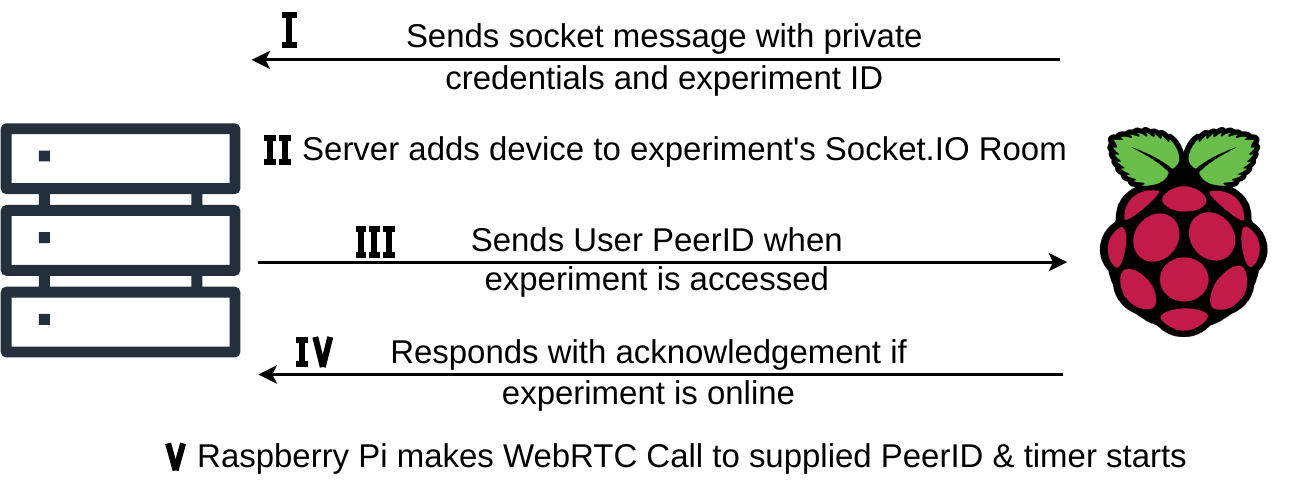}
    \caption{WebSocket communication between Raspberry Pi and server}
    \label{fig:websocket}
    \vspace{-4mm}
\end{figure}

\subsubsection{IoT Component:}

The IoT component comprises the hardware nodes, made of Raspberry Pi, sensors and actuators which facilitate remote experimentation. It also includes the interaction of Raspberry Pi with the experiment frontend for WebRTC video streaming, the Blynk cloud using its proprietary protocol and the WebSocket communication with the backend server. The WebSocket API enables bidirectional communication between a client and a server. The RLabs platform uses Socket.IO\footnote{\url{https://socket.io}}- a library built on top of the WebSocket protocol. It provides additional guarantees like fallback to HTTP long-polling. Once a Raspberry Pi opens the corresponding web-page intended for communications with the setups, a Socket.IO connection is established between it and the server. As depicted in Fig. \ref{fig:websocket}, after initialising the connection, the Raspberry Pi sends a socket message to the server with private credentials and its corresponding experiment's ID to gain authorisation (I). Once a Raspberry Pi is authorised, the server adds the device to a Socket.IO room (II) corresponding to the experiment it is responsible for. These rooms facilitate sending messages in bulk to each Raspberry Pi associated with a given experiment. When a user makes a request to the server to begin an experiment session, post-checking the occupancy and availability of the experiment, the server sends a socket message indicating the same to every Raspberry Pi associated with the experiment (III). This message contains the user's PeerID for WebRTC communication. Once a Raspberry Pi receives a user's credentials, it responds to the server with an acknowledgement message (IV). This informs the server that the experiment is online and unoccupied. Once the Raspberry Pi sends an acknowledgement, it makes a WebRTC call (V) to the user using the supplied PeerID and simultaneously starts a timer for the experiment session. When the allotted time for the session runs out, the Raspberry Pi ends the call and sends a socket message to the server indicating the same. When the server receives a socket message corresponding to a session's end, it updates relevant records and marks the experiment as unoccupied.


The RLabs platform proposed in this paper is a globally deployed web platform, allowing everyone with universal access to use the hardware setups of the above-mentioned experiments remotely on any smart device capable of opening a webpage on a modern browser, requiring no supplementary applications to be downloaded. The platform's frontend is hosted for free on Netlify\footnote{\url{https://www.netlify.com/}} and the backend server and database hosted on Digital Ocean \footnote{\url{https://www.digitalocean.com/}} with a paid Blynk subscription. Moreover, implementing the platform as four separate components makes the platform modular. This makes the RLabs platform easier to scale as both horizontal and vertical scaling can be achieved simultaneously without affecting the other components. Fig. \ref{fig:tech_stack} quickly summarises all the technologies which are used for implementing RLabs. It can be observed that the RLabs platform has been implemented by integrating multiple open-source features, packages, libraries, applications and non-proprietary tools, making the platform easy to replicate. The whole pipeline for remote experimentation built upon these technologies must work properly which reflects the importance of reliability in RLabs. Hence, an automated testing system is proposed to ensure the correct functioning and working of the platform.

\subsection{Automated Testing System}

For a platform allowing users to remotely conduct scientific learning by doing mechanical actuation, the reliability and consistency of results is a major concern. These become even more vital considering that these experiments are utilised for educational and research purposes where accuracy is of prime importance. Hence, after the experiments are deployed, it is crucial to receive real-time updates on the connectivity, availability and working of the experiments to guarantee optimal performance and timely maintenance. These updates are obtained by performing several checks, including the working of hardware components, the functioning of cameras, the stability of network connectivity and the responsiveness of the cloud service. Various strategies for executing these checks manually have been previously proposed. However, the manual execution of these assessments can be time-consuming and inefficient, especially when they need to be performed on a daily basis. 

\begin{figure}[tb]
    \centering
    \includegraphics[width=0.75\columnwidth]{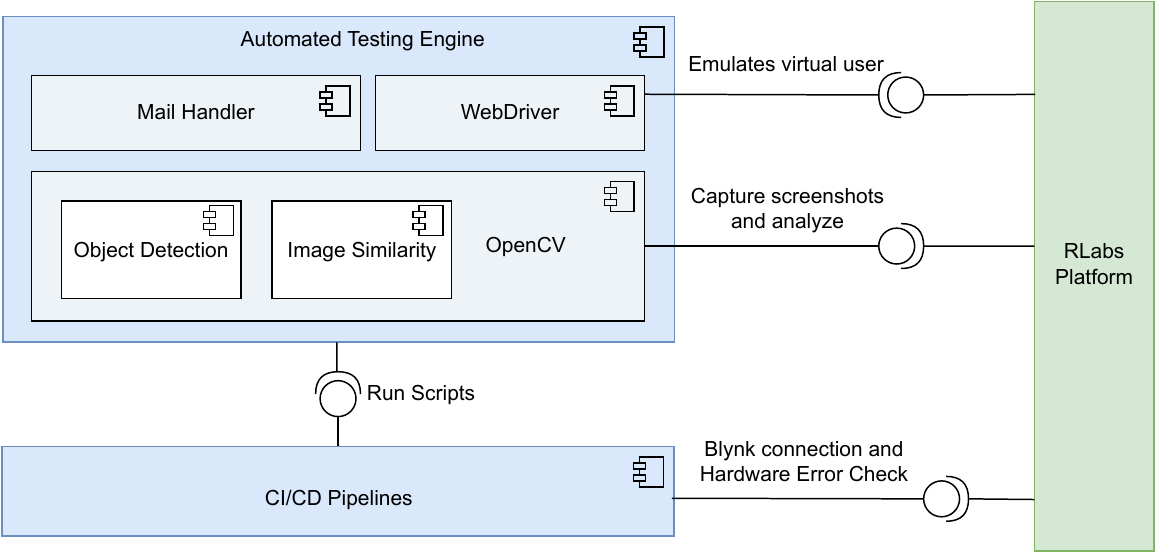}
    \caption{Automated testing component diagram}
    \label{fig:testing_component}
    \vspace{-4mm}
\end{figure}

Automated testing is a process of consistently monitoring the functionality and integrity of experiments showcased on the RLabs platform automatically without any human interference and emerges as a vital solution in this context, facilitating a streamlined, efficient, and hassle-free means of monitoring the experiments. To ensure the maintainability of the platform, including all of its components mentioned above, along with the reliability of the results, an extensive automated testing system has been developed for the RLabs platform. It is designed specifically to identify and help troubleshoot anomalies and irregularities in experimental processes effectively, ensuring the consistency of results. Fig. \ref{fig:testing_component} shows that the \textit{Web Driver, OpenCV scripts and Mail Handler} constitute the testing system, invoked periodically by the Selenium-based automation script, which is deployed on GitHub Actions. The system comprises two major testing elements  - 1) Hardware system checks and 2) Computer-Vision (CV) testing. Both these elements work synchronously for a comprehensive testing system that has been automated using the GitHub Actions workflow, constituting the CI/CD component. Whenever the script detects an error, fault or inaccuracy, the mail handler is used to alert the software administrators regarding it. An elaborate discussion of the elements, components, their implementation and interaction with the RLabs platform is given below.

\subsubsection{Hardware system checks:}
This element of the automated testing system primarily focuses on reporting errors that arise from the scripts running in the background of Raspberry Pi and other microcontrollers. Catch handlers are integrated into the scripts to detect and report errors at various stages. On the hardware side, errors are identified, including those thrown by hardware packages controlling the stepper motor drivers and instances of crossing set thresholds. Hardware assets, including moving components like motors, are also recalibrated after each experiment session to guarantee precision in reporting distances. For instance, the Focal Length experiment is equipped with limit switches that facilitate the recalibration process. Suppose an error is detected in the Raspberry Pi's working or the extended IoT system. In that case, the Raspberry Pi updates a variable on the Blynk Cloud with a pre-decided error code pertaining to the issue or fault that has occurred, which would then be captured by the GitHub workflow described below.

\subsubsection{Computer-Vision testing:}


The hardware system checks described above offer the capability to capture only the errors and faults that might occur in the hardware nodes. As mentioned earlier, the testing system aims to ensure that any anomalies or operational inefficiencies are promptly identified, reported and addressed. For the aim of comprehensive testing, there should be a constant check on the software platform's working along with the hardware's functionality. This is made possible by the OpenCV-based script, which is invoked at regular intervals, typically every eight hours. This comprehensive checking script, although requiring the temporary halting of the experiment, ensures all components are working as intended and uses CV for monitoring mechanical changes such as changes in the position of components, etc. 

\begin{figure}[tb]
    \centering
    \includegraphics[width=\textwidth]{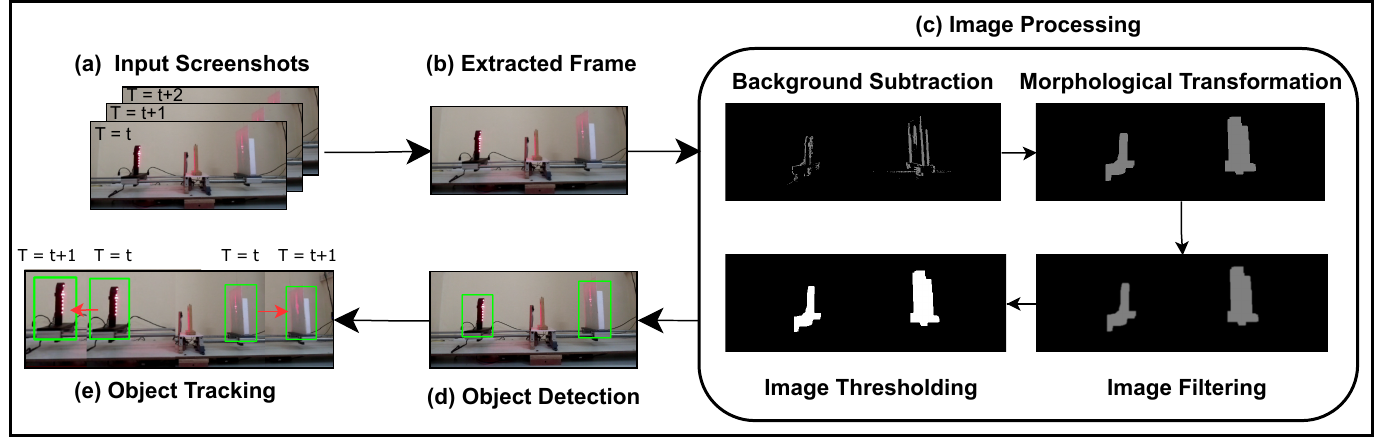}
    \caption{Algorithmic pipeline of the CV script for testing Focal Length experiment.}
    \label{fig:fl_auto}
    \vspace{-4mm}
\end{figure}

Both experiments aim to detect and track moving objects, such as glass rods for the Vanishing Rod experiment and object and screen platforms for the Focal Length experiment. This task can be considered an extension of foreground detection, where moving objects are considered in the foreground. Various methods exist to monitor these changes depending on the specific use case, such as Deep Learning-based tracking or a combination of image processing techniques like background subtraction, optical flow, and morphological transformations. Each approach has its own set of strengths and weaknesses, including computational intensity, real-time performance, accuracy, and robustness in different environmental conditions. The system implemented in this work employs background subtraction and morphological transformations \cite{rtl_cv} due to their lower computational intensity and ability to provide real-time results. Figure \ref{fig:fl_auto} illustrates the algorithmic pipeline of the image processing and CV techniques used in the Focal Length experiment. The techniques used are: \textit{Background Subtraction}, which eliminates the background from screenshots, emphasising only the moving objects and the platforms. Following this, \textit{Morphological Transformation} is applied, particularly the `closing' technique (a combination of dilation and erosion), to reduce image noise and fill small spaces, aiding in more accurate foreground segmentation. Next, \textit{Image Filtering} through median filtering further reduces noise, enhancing image quality. The final step, \textit{Image Thresholding}, involves thresholding the image to enable the drawing of bounding boxes around these foreground objects, completing the process.

Once the bounding boxes are annotated over the object and screen platforms, their locations can be tracked to determine their movements. This helps check if the motors are working correctly and moving to the desired location. In the Vanishing Rod experiment, the system tracks the movement of glass rods, as mentioned above. The experiment's motors are tested by moving the rods up and down, and the data obtained is used to verify their functionality. The beaker's water level is also monitored using Structural Similarity Index Measure (SSIM) \cite{ssim}. This is crucial as the water in the beaker can evaporate over time. In the Focal Length experiment, Fig. \ref{fig:fl_auto} illustrates the usage of the CV pipeline. The system moves the object and screen platforms and tracks the distance travelled by the platforms. Since the platforms move linearly and the total length of the setup is already determined, this fact is used to verify if the platforms have moved to their desired locations, thus verifying the efficiency of the motors.

\begin{figure}[tb]
    \centering
    \includegraphics[width=\columnwidth]{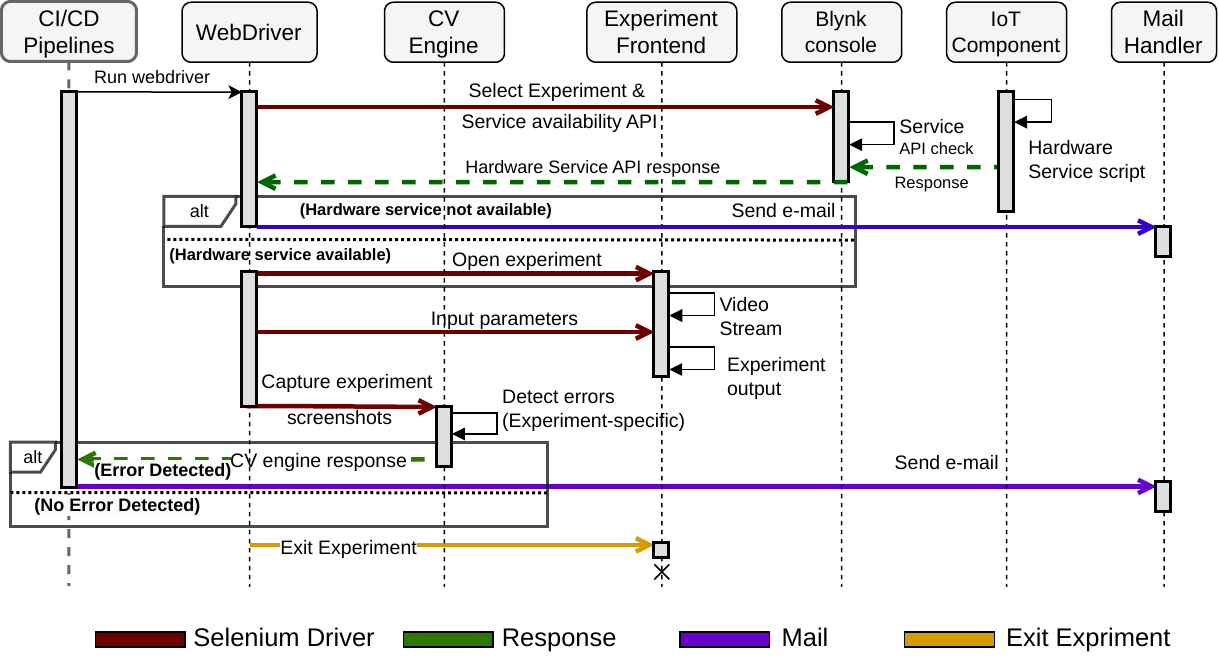}
    \caption{Automated testing sequence diagram}
    \label{fig:testing_sequence}
    \vspace{-4mm}
\end{figure}

\subsection{Execution Flow for Automated Testing}

As mentioned earlier, a Selenium script deployed on GitHub Actions is used to automate the whole testing process. Selenium is a widely used open-source web browser automation software framework, enables interactions with web applications programmatically, such as clicking buttons, filling forms and navigating web pages. The Selenium script emulates a virtual user that logs into the RLabs platform and assesses the experiment's functioning by manipulating the controls on the dashboard. Fig. \ref{fig:testing_sequence} shows that the script deployed on GitHub Actions as a workflow runs periodically (after every eight hours) and starts the Selenium driver to emulate a virtual user that logs into the RLabs platform, selects an experiment to perform and assesses that experiment's functioning by checking the video stream and manipulating the input parameter controls. It initially awaits for the video stream to begin and captures screenshots of the experiment's initial state. Then, the input parameters are given to prompt actuation in the experiment, and multiple screenshots are then captured, which are processed by the CV engine that employs the techniques described earlier to detect if there is any error. Please note that the video stream and the experiment output are obtained on the experiment frontend by following the execution described previously in Fig. \ref{fig:seqeunce_sw}. Hence, the detailed flow for these has been omitted from the sequence diagram for automated testing.

Simultaneously, the GitHub workflow also makes REST API calls to the Blynk Cloud to check if the experiment's hardware is connected to Blynk and if any error code was received from the hardware setups. The workflow receives an appropriate HTTP response. If there is an error raised by the CV engine or the hardware is found to be disconnected or an error code was transmitted from the hardware as part of the system check, an appropriate mail is sent by the mail handler notifying the administrators and the software team regarding the same. In case of perfect remote experimentation, the Selenium driver simply logs out. This sequential flow, as mentioned earlier, is followed every eight hours. As the rules for anomaly detection are highly specific and tailored to the nature of each experiment, particularly the CV techniques, there is a separate testing framework for each experiment with a distinct CV engine. The two scripts for the Vanishing Rod and Focal Length experiment are hosted on the Github Cloud Server and are triggered at predefined intervals by Cron jobs, ensuring regular and automated testing.

\subsection{Additional features of RLabs platform}

The RLabs platform and testing system discussed above together form an exhaustive and comprehensive solution for conducting scientific experiments remotely. However, any system or platform developed for public use also needs to pay close attention to its performance and potential to scale and provide an enhanced user experience. Multiple features and crucial functionalities, focussing on such aspects, have been integrated into the platform, and their elaborate explanation has been given below.

\begin{figure}[tb]
    \centering
    \begin{subfigure}{0.49\textwidth}
        \centering
        \includegraphics[height=5cm,width=\textwidth]{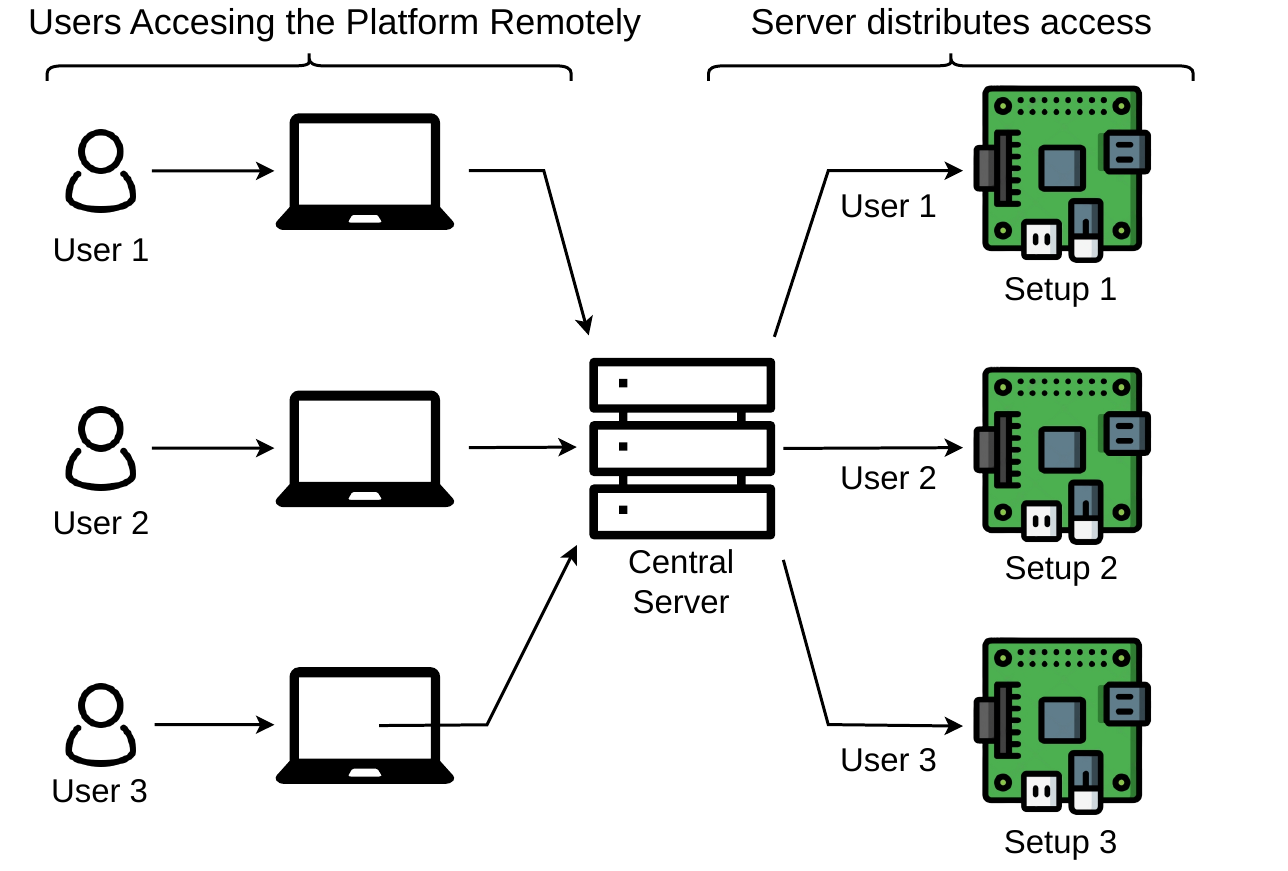}
        \caption{}
        \label{fig:HW_Mult}
    \end{subfigure}
    \hfill
    \begin{subfigure}{0.49\textwidth}
        \centering
        \includegraphics[height=5cm,width=\textwidth]{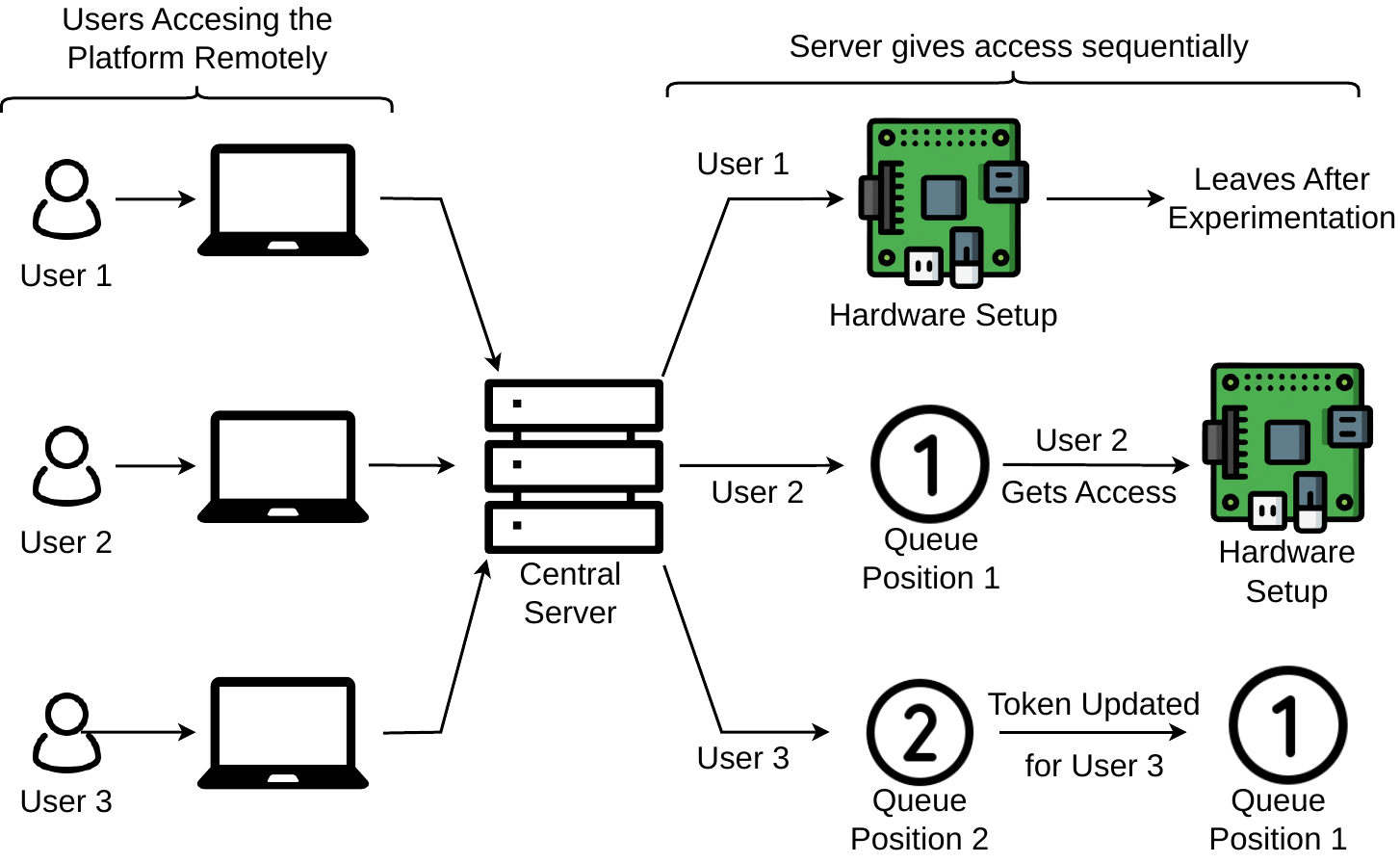}
        \caption{}
        \label{fig:queue_load}
    \end{subfigure}
   \caption{User management in RLabs platform. (a) Hardware multiplexing with 3 hardware nodes of the same experiment. (b) Queue implementation.}
\end{figure}

\subsubsection{User Management on RLabs Platform:}

RLabs experiments involving mechanical actuation present a case of forced single-user access, as multiple users cannot control motor actuation simultaneously. To resolve issues with user accessibility to the experiments and manage the supply and demand for experimentation in such scenarios, methods such as hardware multiplexing, queues and slot booking must be implemented. While hardware multiplexing provides a supply-side solution, queues and slot-booking are software implementations that work on the demand-side of the pipeline enabling remote experimentation.

Hardware multiplexing refers to the platform's ability to accommodate hardware horizontal scaling and handle the increasing nodes of an experiment by redirecting users to different nodes of the same experiment. Horizontal scaling is one of the most efficient ways to provide more user experimentation instances available. Due to the mechanical nature of most of the experiments, there must be single-user access to a particular hardware node. However, by horizontal scaling, multiple users can access multiple experiment instances simultaneously while the platform adheres to the single-user access limitation. This makes hardware instance expansion critical and the ease of integrating multiple instances, once built, into the platform with a few clicks. For adding new instances of the experiments whose dashboard and data presentation formats are pre-designed, the administrators require only the Blynk authentication tokens and the WebRTC initiating experiment room IDs. This horizontal scaling by creating more hardware instances is an example of supply-side management. So, if a user wants to perform a desired experiment and some nodes are already occupied by the other online users, to maintain the single-user access, the student would automatically be given access to another hardware node which is live, running, connected to the platform and currently vacant in terms of usage. The single-user access, along with
hardware multiplexing is depicted in Fig. \ref{fig:HW_Mult}

Queues are a software implementation that is an improvement on the demand-side that forces First-Come, First-Serve access or is referred to as the First-In, First-Out (FIFO) scheduling strategy. Queues are used to give sequential access to the users in the order of when the demand to access was recorded. As depicted in Fig. \ref{fig:queue_load}, if User 1 is already using an experiment and the second user wants to access, they are put in a queue with an incrementing waiting queue token number. When the first user leaves, and the experiment is free to use again, User 2 gets the first chance to experiment, and the token numbers reduce for all the users in the waiting queue. A combination of hardware multiplexing and queues can handle a large load efficiently. Slot Bookings are another widespread demand-side management software implementation. This refers to the platform's ability to distribute the user demand of accessing hardware nodes over slots of predecided time intervals. Users can reserve and book an experiment for a particular time duration to perform remote experimentation. Such slot bookings can also be used to resolve high demands on such remote lab solutions.

\subsubsection{Multiple video streams for the same experiment:}

The software platform must also be compatible to support multiple camera streams for a single experiment. The Focal Length experiment explained above provides two views of the experiment's hardware node - a closeup view of the image on the screen and a side-view of the entire experiment's setup. Using a combination of WebRTC and Socket.IO-based streaming, the server adds all the Raspberry Pis pertaining to a single hardware instance to a dedicated Socket.IO room post-authorisation. When a user tries accessing a hardware instance, server post-user authentication sends a socket message with the user PeerID to each Raspberry Pi in the dedicated room of the desired experiment, after which each Raspberry Pi initiates a WebRTC call individually. This seemingly complex pipeline allows RLabs to have numerous cameras if required by an experiment.

%% file: source_results.tex
This section presents the performance of the proposed RLabs system based on the NFAs proposed earlier. This will include cost tables for building the experiments to demonstrate their affordability, assembly steps for the miniaturised setups that make the experiments portable, compatibility of the RLabs system with different hardware boards, IoT platforms and cameras, availability of the deployed experiments, and an analysis of the user survey collected. 

\begin{table}[tb]
\centering
\caption{Cost tables for procuring the components of the two use case experiments}
\begin{subtable}[t]{.5\textwidth}
  \centering
  \caption{Costs for lab-scale Vanishing Rod setup}
  \label{tab:lab_vr_cost}
  \begin{tabular}{|lll|}
    \hline
    \multicolumn{3}{|c|}{\textbf{Lab-Scale VR}}  \\ \hline
    \multicolumn{1}{|l|}{\textbf{Item}} & \multicolumn{1}{l|}{\textbf{Qty}} & \textbf{Cost} \\ \hline
    Raspberry Pi 3B+ & 1         & 5000          \\
    28BYJ-48           & 2         & 160           \\
    ULN2003            & 2         & 80            \\
    Body Frame         & 1         & 500           \\
    Camera             & 1         & 500           \\
    Glass Rod          & 2         & 60            \\
    Beaker             & 2         & 400           \\
    Misc               & 1         & 1000          \\
    &&\\
    \textbf{Total}     & \textbf{} &  \multicolumn{1}{c|}{\textbf{\begin{tabular}[c]{@{}c@{}}7700 INR\\ 92 USD\end{tabular}}} \\ \hline
  \end{tabular}
\end{subtable}%
\begin{subtable}[t]{.5\textwidth}
  \centering
  \caption{Costs for miniaturised Vanishing Rod setup}
  \begin{tabular}{|lll|}
    \hline
    \multicolumn{3}{|c|}{\textbf{Miniature VR}} \\
    \hline
    \multicolumn{1}{|l|}{\textbf{Item}} & \multicolumn{1}{l|}{\textbf{Qty}} & \textbf{Cost} \\
    \hline
    Raspberry Pi Zero 2W & 1 & 1650 \\
    Body Frame & 1 & 1000 \\
    Camera & 1 & 500 \\
    Fabricated PCB & 1 & 300 \\
    28BYJ-48 & 1 & 80 \\
    ULN2003 & 1 & 40 \\
    ESP32 & 1 & 400 \\
    Glass Rod & 2 & 30 \\
    Beaker & 2 & 100 \\
    Misc & 1 & 400 \\
    &&\\
    \textbf{Total} & &  \multicolumn{1}{c|}{\textbf{\begin{tabular}[c]{@{}c@{}}4500 INR\\ 54 USD\end{tabular}}} \\
    \hline
  \end{tabular}
  \label{tab:mini_vr_cost}
\end{subtable}
\hspace{4mm}
\begin{subtable}[t]{.5\textwidth}
  \centering
  \caption{Costs for lab-scale Focal Length setup}
  \begin{tabular}{|ccc|}
    \hline
    \multicolumn{3}{|c|}{\textbf{Lab-Scale FL}} \\
    \hline
    \multicolumn{1}{|c|}{\textbf{Item}} & \multicolumn{1}{c|}{\textbf{Qty}} & \multicolumn{1}{c|}{\textbf{Cost}} \\
    \hline
    Raspberry Pi 3B+ & 1 & 5000 \\
    NEMA 17 & 2 & 1700 \\
    A4988 & 2 & 300 \\
    USB camera & 1 & 1200 \\
    RaspiCam & 1 & 500 \\
    8mm Axle & 4 & 1200 \\
    Axle support & 8 & 400 \\
    Screw Rod + Nut & 2 & 650 \\
    Shaft coupler & 2 & 150 \\
    Bearing & 2 & 200 \\
    Slider & 4 & 800 \\
    3D printed parts & 1 & 300 \\
    Wooden Planks & 1 & 600 \\
    Misc & 1 & 1500 \\
    &&\\
    \textbf{Total} & &  \multicolumn{1}{c|}{\textbf{\begin{tabular}[c]{@{}c@{}}14500 INR\\ 174 USD\end{tabular}}} \\
    \hline
  \end{tabular}
  \label{tab:lab_fl_cost}
\end{subtable}%
\begin{subtable}[t]{.5\textwidth}
  \centering
  \caption{Costs for miniaturised Focal Length setup}
  \begin{tabular}{|ccc|}
    \hline
    \multicolumn{3}{|c|}{\textbf{Miniature FL}} \\
    \hline
    \multicolumn{1}{|c|}{\textbf{Item}} & \multicolumn{1}{c|}{\textbf{Qty}} & \textbf{Cost} \\
    \hline
    Raspberry Pi zero 2W & 2 & 3300 \\
    NEMA 17 & 2 & 1700 \\
    A4988 & 2 & 300 \\
    Camera & 2 & 1000 \\
    Support Wheels & 4 & 400 \\
    Aluminium Profile & 2 & 800 \\
    Timing Belt & 2 & 200 \\
    Pulleys & 2 & 150 \\
    Idler & 2 & 150 \\
    3D printed parts & 1 & 1500 \\
    Fabricated PCB & 1 & 500 \\
    Misc & 1 & 1000 \\
    &&\\
    \textbf{Total} & &  \multicolumn{1}{c|}{\textbf{\begin{tabular}[c]{@{}c@{}}11000 INR\\ 132 USD\end{tabular}}} \\
    \hline
  \end{tabular}
  \label{tab:mini_fl_cost}
\end{subtable}
\label{tab:cost_tables}
\end{table}

\subsection{Low-cost Hardware Setups}

Table \ref{tab:cost_tables} shows the total material costs for building the lab-scale and miniaturised setups for the two use case experiments in INR. It can be observed that the total cost of the lab-scale Vanishing Rod and Focal Length experiments are 7700 INR (92 USD\footnote{1 USD is approximately 83 INR as of December 2023}) and 14500 INR (174 USD), respectively. For the miniaturised Vanishing Rod and Focal Length experiments, the total cost of the setups was 4500 INR (54 USD) and 11000 INR (132 USD). There is a 41 \% and 24 \% reduction in costs for making the miniaturised setups for Vanishing Rod and Focal Length experiments, respectively. Firstly, the setups are low-cost and can be attributed to using single-board computers, which has brought down the costs as individual servers or PCs are not required for hosting the experiments. Raspberry Pi 3B+ (5000 INR) is the costliest item (33 \% and 66 \%) in the lab-scale setups. This cost can be reduced by replacing it with cheaper boards like Raspberry Pi Zero 2 W (1650 INR), significantly reducing the costs observed in the miniaturised setups.

\begin{table}[tb]
\caption{Comparison of lab-scale and miniaturised experimental setups}
\centering

\begin{tabular}{|c|cc|cc|}
\hline
\multicolumn{1}{|l|}{\multirow{2}{*}{Type   of Setup}} & \multicolumn{2}{c|}{Lab-scale setup}          & \multicolumn{2}{c|}{Miniaturised setup}                \\ \cline{2-5} 
\multicolumn{1}{|l|}{}                                 & \multicolumn{1}{c|}{VR}       & FL        & \multicolumn{1}{c|}{VR}       & FL       \\ \hline
Mass (KG)                                              & \multicolumn{1}{c|}{1.8}      &      \multicolumn{1}{c|}{6.2}     & \multicolumn{1}{c|}{0.33}     &     \multicolumn{1}{c|}{2.2}     \\ \hline
Volume ($cm^3$)                                            & \multicolumn{1}{c|}{30$\times$20$\times$32} & 25$\times$150$\times$20 & \multicolumn{1}{c|}{10$\times$10$\times$16} & 10$\times$85$\times$20 \\ \hline

\end{tabular}
\label{tab:lab_vs_mini}
\end{table}
\begin{figure*}[tb]
\centering
\includegraphics[width=\textwidth]{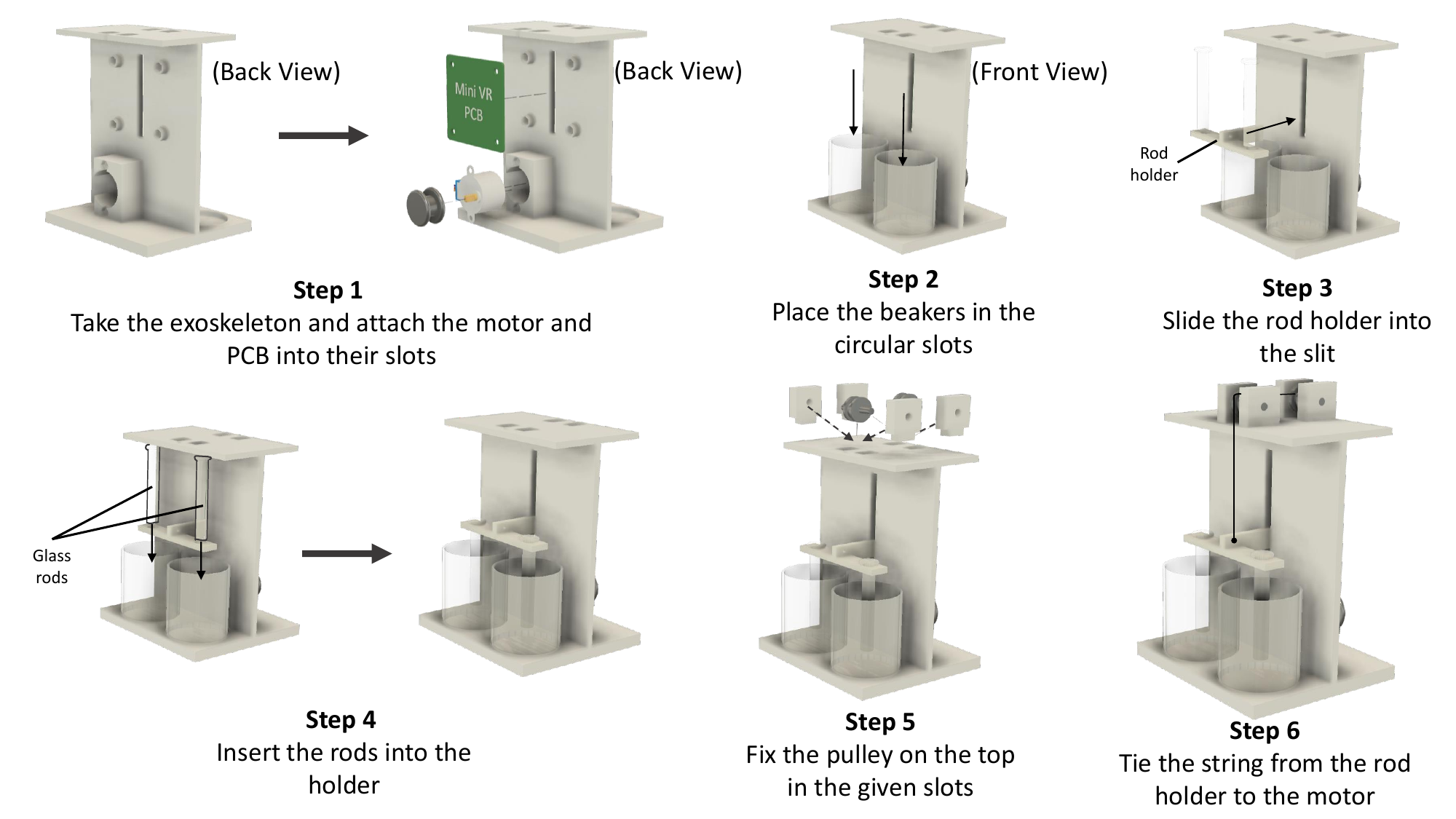}
\setlength{\belowcaptionskip}{5pt}
\caption{Assembly of the miniature Vanishing Rod setup}
\label{fig:miniVR_assembly}
\end{figure*}

\begin{figure*}[tb]
    \centering
    \includegraphics[width=\textwidth]{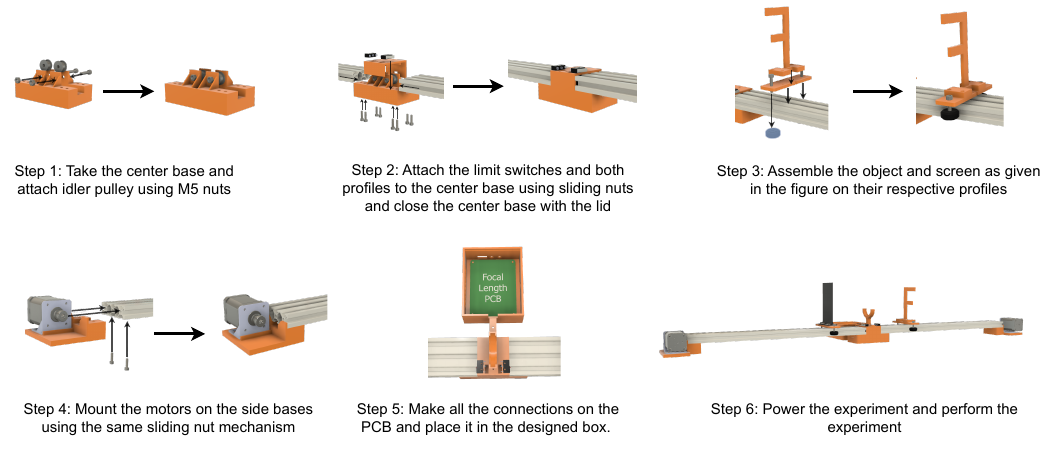}
    \caption{Assembly of the miniature Focal Length Setup}
    \label{fig:miniFL_assembly}
    \vspace{-4mm}
\end{figure*}

\subsection{Portable Hardware Setups}

To showcase the portability of the built experiments, the mass and volume of the experiments are reported along with their step-by-step assembly. Table \ref{tab:lab_vs_mini} presents a comparative analysis of the lab-scale and miniaturised setups, focusing on mass (excluding the beakers filled with oil and water) and volume. There has been a 5.4 and 12-times reduction in the weight and volume of the Vanishing Rod experiment from the lab-scale to the miniaturised setup. Similarly, in the case of the Focal Length experiment, the weight and volume were reduced by a factor of 2.8 and 4.4, respectively. 

\par Figs. \ref{fig:miniVR_assembly} and \ref{fig:miniFL_assembly} show the step-by-step assembly of the two use case experiments from their components in a modular fashion. The setups are designed using various techniques to make their assembly easy. The parts have different shapes that fit together, making it intuitive for the user to join them. This also eliminates the use of screws and bolts to some extent. For example, in the Vanishing Rod setup, the exoskeleton has a specially designed mount for the motor that exactly fits over the motor and holds it without any screws. Similarly, sliding nuts and the aluminium profile create a slide and lock mechanism for the Focal Length setup. The sliding nuts easily slide into the profile slots and can be tightened using a bolt that holds the centre base and other parts against the profile. This system overcomes the need to drill holes into the profile, making it easier for an individual to assemble the setup. Both setups are entirely made of off-the-shelf parts along with 3D-printed modules that are usually available in the local hardware and electronics stores. The combination of modularity and miniaturisation renders the setups portable, similar to IKEA\footnote{\url{https://www.ikea.com/}} products, enabling users to assemble and disassemble them easily. This portability facilitates the shipment of setups to various locations worldwide, including remote rural areas, thereby extending their accessibility.

\subsection{Compatibility with Different Hardware Boards, Cameras and IoT Platforms}

In order to ensure broader compatibility, the software architecture is designed to focus on supporting a wide range of hardware boards. This is shown using different boards (such as Raspberry Pi and ESP32) to implement use case experiments. The boards used must be able to connect online using any IoT platform that facilitates data access via APIs. This allows for the accommodation of various hardware boards. However, there can be boards that do not support internet connectivity. Those boards can then be coupled with an ESP8266 or ESP32, which provides smooth data transfer from the experiment to the dashboard at an economical price point, starting from as low as 400 INR ($\sim$5 USD).

\par If a board does not support live-streaming from a camera, a Raspberry Pi Zero 2 W\footnote{\url{https://www.raspberrypi.com/products/raspberry-pi-zero-2-w/}}, which costs around 1600 INR ($\sim$20 USD), can be used alongside standard RaspiCams solely for streaming purposes. For experiments that use a Raspberry Pi, any IP camera or USB camera compatible with and recognised by a Raspberry Pi can be used for streaming. Standalone IP cameras can be easily set up by users with minor configurations, including setting up a static IP for the camera and port forwarding to grant global access. Table \ref{tab:cam_delay} shows the observed delay in live-streams across different devices. The WebRTC facilitates the streaming in Raspberry Pis, while the stream from the TP-Link Tapo C100 IP Camera\footnote{\url{https://www.tapo.com/in/product/smart-camera/tapo-c100/}} is accessed using Real Time Streaming Protocol (RTSP). The Raspberry Pi Zero W\footnote{\url{https://www.raspberrypi.com/products/raspberry-pi-zero-w/}} was evaluated with the streaming script as well. However, it delivered a poor-quality stream and exhibited significant lag, with delays exceeding 3 seconds. Alternatively, the ESP32 Cam module offers streaming capabilities, requiring minor tweaks to support WebRTC \cite{esp_webrtc}. However, it is essential to note that the authors have not officially tested this configuration.

\par Once the hardware experiments are built, the next step is their integration with the RLabs platform. The Blynk cloud service, an IoT platform currently supporting HTTP(S), is used for this. However, the RLabs platform can handle other protocols from various IoT platforms, provided they allow reading and modification of variable states essential for designing experiments. To test this, the dashboard for the Focal Length experiment is created and tested on two IoT platforms: Blynk using HTTPS and Thingspeak\footnote{\url{https://thingspeak.com/}} via MQTT. 
Other notable IoT platforms include Arduino IoT Cloud\footnote{\url{https://docs.arduino.cc/arduino-cloud/}}, and AWS IoT Core\footnote{\url{https://aws.amazon.com/iot-core/}}. These platforms extend their support to multiple hardware boards, including Raspberry Pi and ESP boards, through specialised libraries and packages, making integration even simpler.

\begin{table}[tb]
    \centering
      \caption{Latency in live-streaming of video across various cameras}

\begin{tabular}{ll}
\hline
\multicolumn{1}{|l|}{Camera} & \multicolumn{1}{l|}{\begin{tabular}[c]{@{}l@{}}Delay (in sec)\end{tabular}} \\ \hline
\multicolumn{1}{|l|}{RaspiCam + Raspberry Pi 3B+}      & \multicolumn{1}{l|}{0.35} \\ \hline
\multicolumn{1}{|l|}{RaspiCam + Raspberry Pi Zero 2 W} & \multicolumn{1}{l|}{0.86} \\ \hline   
\multicolumn{1}{|l|}{TP-Link Tapo C100 IP cam}               & \multicolumn{1}{l|}{2.01} \\ \hline

\end{tabular}%
  \label{tab:cam_delay}

\end{table}

    \begin{figure}[tb]
        \centering
        \includegraphics[width=0.65\textwidth]{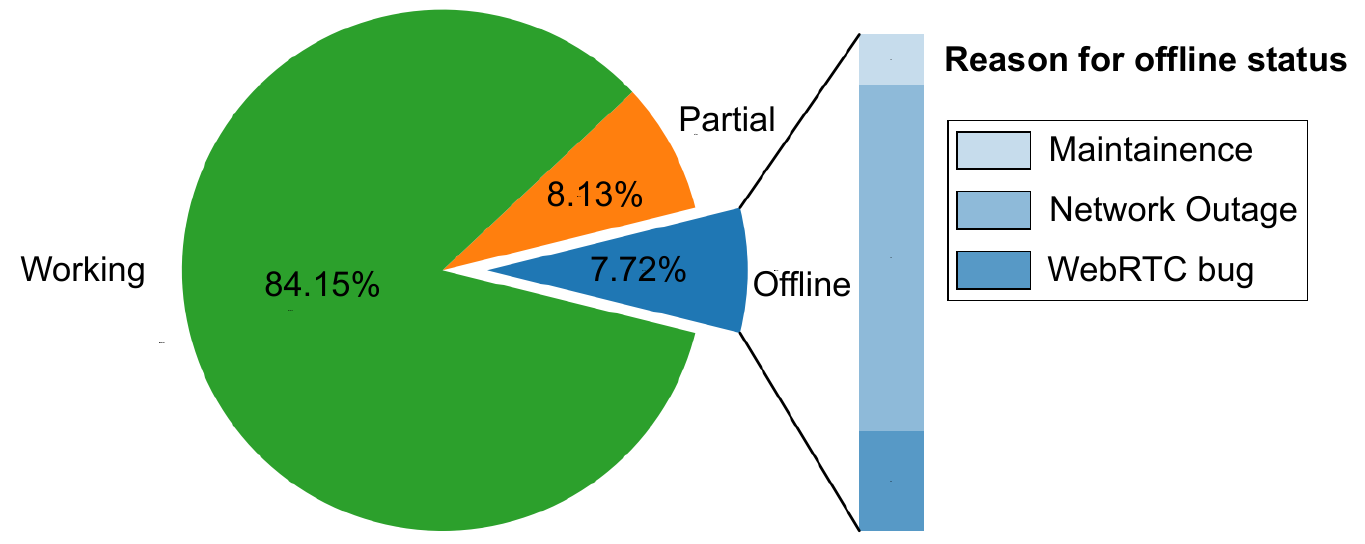}
        \vspace{-3mm}
        \caption{Availability of the experiments}
        \label{fig:avail}
        \vspace{-4mm}
    \end{figure}

\subsection{Availability of the Experiments}

Fig. \ref{fig:avail} displays the operational status of the experiments recorded by the aforementioned automated testing suite. Daily, over four months from July to October 2023, the operational status of the lab-scale experiments was logged.

The findings have been segmented into three distinct categories:

\begin{enumerate}
\item Online: Indicates that both the experiment's hardware and the live-stream are functioning optimally without disruptions throughout all the readings taken for the day.
\item Partial: Signifies either a malfunction in the live-stream service or an issue with the experiment's hardware at least once during the day.
\item Offline: Specifies that the experiment's hardware is entirely non-operational for the day.
\end{enumerate}

Out of the 123 days (4 months), the Vanishing Rod experiment was online for 106 days, partially working for eight days, and entirely offline for nine days. The Focal Length experiment was online for 101 days, partially working for 12 days and entirely offline for 10 days. In July, the experiments generally ran smoothly, with a few exceptions, such as when an issue with the implementation of WebRTC caused the experiments to appear offline. This issue was identified and resolved within a few days. The experiments then operated without further problems. However, towards the latter part of August and early September, the labs experienced significant downtime, primarily due to a severe disruption in the campus network where the experiments were hosted. This outage persisted for nearly seven days before being rectified. There were a few instances of experiments disconnecting from the Wi-Fi network and failing to reconnect. The exact cause has yet to be determined, but it may be related to Wi-Fi router-specific issues \cite{wifi_connect}. This record-keeping has provided invaluable insights into the system's reliability, serving as a critical measure of its performance. By understanding these patterns, proactive steps can be taken to enhance the system's reliability.


\begin{table}[tb]
    \caption{Average Scores and Responses of User Feedback on Remote Labs}
    \centering

    \begin{tabular}{|c|c|}
    \hline
    \textbf{Questions} & \textbf{Average Score} \\
    \hline
    \hline
    The remote lab significantly helped in my learning process & 4.09 \\
    \cline{1-2}
     I felt engaged while using the remote lab & 4.27 \\
    \cline{1-2}
    Using the remote lab felt similar to using a physical lab & 4.24 \\
    \hline
    The remote lab is easy to navigate and use & 4.64 \\
    \cline{1-2}
    The quality of the live video stream was excellent & 4.27 \\
    \hline
    The experiments in the live-stream were highly responsive & 4.40 \\
    \cline{1-2}
    The UI/dashboard and controls for each experiment were responsive and intuitive & 4.49 \\
    \cline{1-2}
    \textbf{Overall Average Score} & \textbf{4.34} \\
    \hline
    \end{tabular}
    \label{tab:user-feedback}
    \vspace{-3mm}
\end{table}

\subsection{User Feedback}

A user survey was conducted, and feedback from a group of forty-five grade-9 students at Shikhar Educare, Amravati, Maharashtra, India was collected. The two use case experiments are part of their curriculum. Table \ref{tab:user-feedback} displays the specific questions posed and the average scores for each. The survey included standard questions \cite{feedback, feedback_2} that focused on pedagogy, usability, and learning outcomes. Additionally, users were asked to rate aspects such as the responsiveness of the experiments, live-streaming quality, and the intuitiveness of the software platform, among others. Responses were provided using a Likert scale, where a score of 1 represented strong disagreement, and 5 indicated strong agreement.

The user feedback on the remote lab solution, as shown in Table \ref{tab:user-feedback}, indicates a positive user experience, with high scores in areas related to usability, clarity of streamed results, and content quality, suggesting that the platform is both user-friendly and educationally effective. The scores ranged from 4.09 to 4.64 on a 5-point scale, with an overall average of 4.34. Students felt the remote lab was easy to navigate and use, receiving the highest score. Most users also found the experiment parameters and feedback form sufficient, reinforcing the solution's comprehensiveness and effectiveness. The users have requested to add other experiments that could complement their understanding in their comments collected at the end. The results overall suggest that the remote lab solution successfully meets user expectations in various vital aspects. However, minor improvements, like adding experiments from different subjects, could enhance the experience further.